\documentclass[12pt]{iopart}
\usepackage{color}

\newcommand{\D}{{\rm d}}
\usepackage[english]{babel}
\usepackage{graphicx}
\usepackage{cite}
\usepackage{subfigure}
\usepackage{iopams}
\usepackage{amssymb}
\usepackage{amstext}

\newcommand{\0}{\underline{\;\,}\,}
\newcommand{\1}{\underline{\bullet}\,}
\newcommand{\binom}[2]{{#1 \choose #2}}

\begin{document}
\title{Analytical study  of an exclusive genetic switch}
\author{J. Venegas-Ortiz and M. R. Evans}
\address{SUPA, School of Physics and Astronomy, University of
Edinburgh, Mayfield Road, Edinburgh EH9 3JZ, United Kingdom}
\eads{\mailto{J.Venegas-Ortiz@sms.ed.ac.uk}, \mailto{m.evans@ed.ac.uk}}

\date{\today, last edited by MRE}

\begin{abstract}
The nonequilibrium stationary state of an exclusive genetic switch
is considered. The model comprises  two competing species and a single
binding site which, when bound to by a protein of one species, causes
the other species to be repressed. The model may be thought of as a
minimal model of the power struggle between two competing parties. 
Exact solutions are given for the limits
of vanishing binding/unbinding rates and
infinite binding/unbinding rates.
A mean field theory is introduced   which is exact
in the limit of  
vanishing binding/unbinding rates.
The mean field theory and numerical simulations reveal that generically bistability occurs and
the system is in a symmetry broken state.  An exact perturbative
solution which in principle allows the nonequilibrium stationary state
to be computed is also developed and computed to first and second order.

\end{abstract}
\maketitle

\section{Introduction}
Genetic networks are interacting, many-component systems of genes, RNA and
proteins, that control the functions of living cells
\cite{bio:alon,WW05,visco:binary}.
They exhibit complex,
nonlinear behaviour due to the feedback between the expression of different
genes.  At a simplistic level of description the gene sequence, through
transcription by RNA and translation by mRNA, leads to the production of
proteins. These proteins may themselves act as transcription factors which
may switch on or off the activity of other genes \cite{swi:ptashne}.  A particularly simple
example of a genetic network is a toggle  switch formed from pairs of genes
that mutually repress each other's expressions \cite{gen:cherry}.
Such switches   serve as
microscopic  models  for bistable and
 oscillatory states \cite{WW04,BB08,VAE08,gen:lipsh}
and importantly  may be synthesized \cite{gen:gardner}.

Genetic switches may exhibit bistability where there are two possible
dynamically stable long-lived states for the switch.  There has been
considerable interest in how bistability may be maintained and how it is
effected by stochastic fluctuations due to small numbers of proteins and
intrinsic noise \cite{sma:mcad,sma:arkin,sma:beck}.  In particular the switching 
time between the two states has been measured and numerical techniques have been 
devised to study the switching time within theoretical 
models \cite{KE01,AWW05,gen:loinger,BB08}.

In this paper we consider the simplest toggle switch 
introduced by  Warren and P.R. ten Wolde \cite{WW04}
which we will refer to as the Exclusive Switch (this model is referred to in
\cite{BB08} as the exclusive switch without cooperative binding). 
The model comprises two genes
labelled 1,2 each leading to the production of proteins $X_1$ and $X_2$.
These proteins also degrade stochastically. There is a single mutual binding
site to which either an $X_1$ or $X_2$ may bind and when bound repress the
production of the other protein.  Thus when an $X_1$ is bound, the $X_1$
population fluctuates around some steady state value determined by the
balance of production and degradation while the $X_2$ population degrades
towards zero.  

This switch may be considered more generally as a minimal
model for the power struggle between two competing parties.  This is
illustrated by the following charicature of ``mob dynamics''. There are two
competing parties or gangs of individuals and room for only one individual
to wield absolute power.  When an individual of one party is in power his
own party membership may grow but the other party membership dwindles.
Random influences imply that the control of power is occasionally lost and
power may be seized by any individual.  Thus in a temporary power vacuum the
membership of the minority party will increase and there is a small chance
that a member of this party will seize power and that the minority will
eventually become the majority.

In the context of statistical physics the Exclusive Switch is an
example of a nonequilibrium system. This is because the microscopic
stochastic dynamics do not obey detailed balance. For example when
an $X_1$ protein is bound the degradation of an $X_2$ protein is
irreversible. The structure of nonequlibrium stationary states has
been of considerable interest and is generally characterised by the
existence of probability currents in the stationary state (which do
not exist in equilibrium stationary states due to the presence of
detailed balance)\cite{Mukamel00}.  For example it has been shown that spontaneous symmetry breaking may occur in non-equilibrium systems under
conditions where it is precluded from their equilibrium counterparts
e.g. in one spatial dimension \cite{EFGM95a,GLEMSS95}.

The bistability exhibited in the Exclusive Switch may be thought of as
symmetry breaking where although the microscopic dynamics is symmetric
between the two proteins, the stationary state comprises two possible
long-lived dynamical states in which the symmetry is broken and one
protein dominates. In an equilibrium system the switching time between
the two symmetry-broken states may be estimated by the Arrhenius law
$\tau \sim \exp \beta \Delta F$ where the free energy barrier $\Delta
F$ is extensive in the system size.  For a nonequilibrium system on
the other hand the free energy or indeed the stationary state is not
known a priori and one is required to construct the stationary state
on a model by model basis.
For  genetic switches there has been recent interest in
developing analytical approaches to describe the stationary states
\cite{SS08,SQ11}.

In the present work we study analytically the nonequilibrium
stationary state of the Exclusive Switch. Previous analytical studies have
concentrated on systems with only one gene
\cite{SS08,onegene:karma,onegene:hornos}. Our aim is to understand
whether symmetry breaking occurs and, if so, the nature of the
symmetry broken state.  To this end we develop two analytical
approaches.  First we construct a mean field theory. Then we develop a
perturbative approach that in principle allows the nonequilibrium
stationary state to be computed exactly and 
we present analytical results to first
order. A complementary approach to this system has been developed in
\cite{swi:ohkubo}, where an approximation scheme based on effective
interactions is used.

The paper is organised as follows.  In section 2 we define the
Exclusive Switch model and write down the system of master equations
that describe the system.  We also present some numerical simulations
which illustrate the nature of the symmetry breaking and consider
exactly solvable limits.  In section 3 we present a mean field theory
and compare to stochastic simulations. In section 4 we develop a
perturbative approach, compute the results to first order and compare
with simulation results.  Conclusions are drawn in section 5.

\section{Model Definition}

The state of the system is defined by:
the number of free proteins of type 1, $N_1$ ;
the number of free proteins of type 2, $N_2$, 
and the  state of the switch, $S$, which takes value 0 if
no protein is bound, value 1 if
a protein of type 1 is bound,
and value 2 if a protein of type 2 is bound.

The stochastic dynamical processes are as follows: a protein degrades
(leaves the system) with rate $d$; when the switch state is 0 proteins
of both type 1 and 2 are produced with rate $g$; when the switch state
is 1 proteins of type 1 produced with rate $g$ and when the switch
state is 2 proteins of type 2 produced with rate $g$; if the switch
state is 0 a protein binds with rate $b$ and when binding occurs the
switch state changes to the type of bound protein and the number of
free proteins is reduced by 1; a bound protein unbinds with rate $u$
and when unbinding occurs the switch state changes to 0 and the
number of free proteins is increased by 1.

We shall consider the joint probabilities
 $P_S(N_1,N_2)$
 of the protein numbers
$N_1$, $N_2$, switch state $S$.

\subsection{Master equation}
Following the stochastic dynamical processes described above
the system of master equations that defines the model 
can be written as  follows
\begin{eqnarray}
\frac{\partial P_0}{\partial t}(N_1,N_2) &=&g[P_0(N_1-1,N_2)+P_0(N_1,N_2-1)-2P_0(N_1,N_2)]\nonumber \\
&+&d[(N_1+1)P_0(N_1+1,N_2)+(N_2+1)P_0(N_1,N_2+1)\nonumber\\
&&-(N_1+N_2)P_0(N_1,N_2)]-b(N_1+N_2)P_0(N_1,N_2)\nonumber \\
&+&u[P_1(N_1-1,N_2)+P_2(N_1,N_2-1)]\label{master0} \\
\frac{\partial P_1}{\partial t}(N_1,N_2) &=&g[P_1(N_1-1,N_2)-P_1(N_1,N_2)]\nonumber \\
&+&d[(N_1+1)P_1(N_1+1,N_2)+(N_2+1)P_1(N_1,N_2+1)\nonumber\\
&&-(N_1+N_2)P_1(N_1,N_2)]\nonumber\\
& +&b(N_1+1)P_0(N_1+1,N_2)-uP_1(N_1,N_2)
\label{master1} \\
\frac{\partial P_2}{\partial t}(N_1,N_2) &=&g[P_2(N_1,N_2-1)-P_2(N_1,N_2)]\nonumber\\
&+&d[(N_1+1)P_2(N_1+1,N_2)+(N_2+1)P_2(N_1,N_2+1)\nonumber\\
&&-(N_1+N_2)P_2(N_1,N_2)]\nonumber\\
& +&b(N_2+1)P_0(N_1,N_2+1)-uP_2(N_1,N_2)
\label{master2}
\end{eqnarray}
where $g$ is the generation rate of a protein (when the generation is
not suppressed by the switch state), $d$ is the degeneration rate of a
single protein, $b$ is the binding rate of a single protein and $u$ is
the unbinding rate of the bound protein.  The whole problem is clearly
symmetric with respect to the variables $1$ and $2$.  Also note that,
the degeneration term is the same for the three equations, since it
does not depend on the value of $S$.

\subsection{Nature of symmetry breaking}
In order to illustrate the qualitative behaviour, we first present
stochastic simulations of the Exclusive Switch. These simulations are 
performed with a Gillespie algorithm \cite{sm:gillespie,simgs:schultz}, in which the reactions described
in the model are given a certain probability, depending on the state of the 
system and the value of the parameters. These probabilities are used to 
determine stochastically which reaction is going to happen next, and when it
will happen. The time that the system spends in a given state $N_1,N_2,S$ is 
normalized to obtain the probability distributions. The reference values for
the simulations are the typical values for bacteria such as Escherichia coli 
\cite{bio:alon} 
\begin{equation}
 g=0.05, \quad d=0.005, \quad b=0.1, \quad u= 0.005
\label{ecv}
\end{equation}

In the system there is a clear symmetry between the two proteins
species, since they undergo the same microscopic reactions.  However,
at any  given time the
system is typically dominated by one of the proteins; thus the symmetry is
broken.  The reason for this is that, once a protein, e.g. of type
$X_1$, binds to the promoter site, proteins $X_2$ start disappearing,
while the number of $X_1$ fluctuates around a steady value. That means
that when the bound protein unbinds (as it will do eventually due
to stochasticity of the system), it is much more probable for proteins
$X_1$ to bind to the promoter site again, since there are more of them.
At the same time  it is  more difficult for proteins $X_2$ to bind to the promoter site. However the $X_2$ will not disappear permanently from the
system (there is no absorbing state), and will be produced again the
moment the bound protein $X_1$ unbinds. Thus, with a small
probability, proteins $X_2$ will be able to bind again to the promoter
site, and become the dominant species, as proteins $X_1$ start to
degenerate.  This means that there are two symmetry-broken states, in
which one species is much more abundant than the other.

\begin{figure}[htb]
 \centering
 \begin{minipage}[c]{0.3\linewidth}
    \begin{center}
        \includegraphics[width=\linewidth]{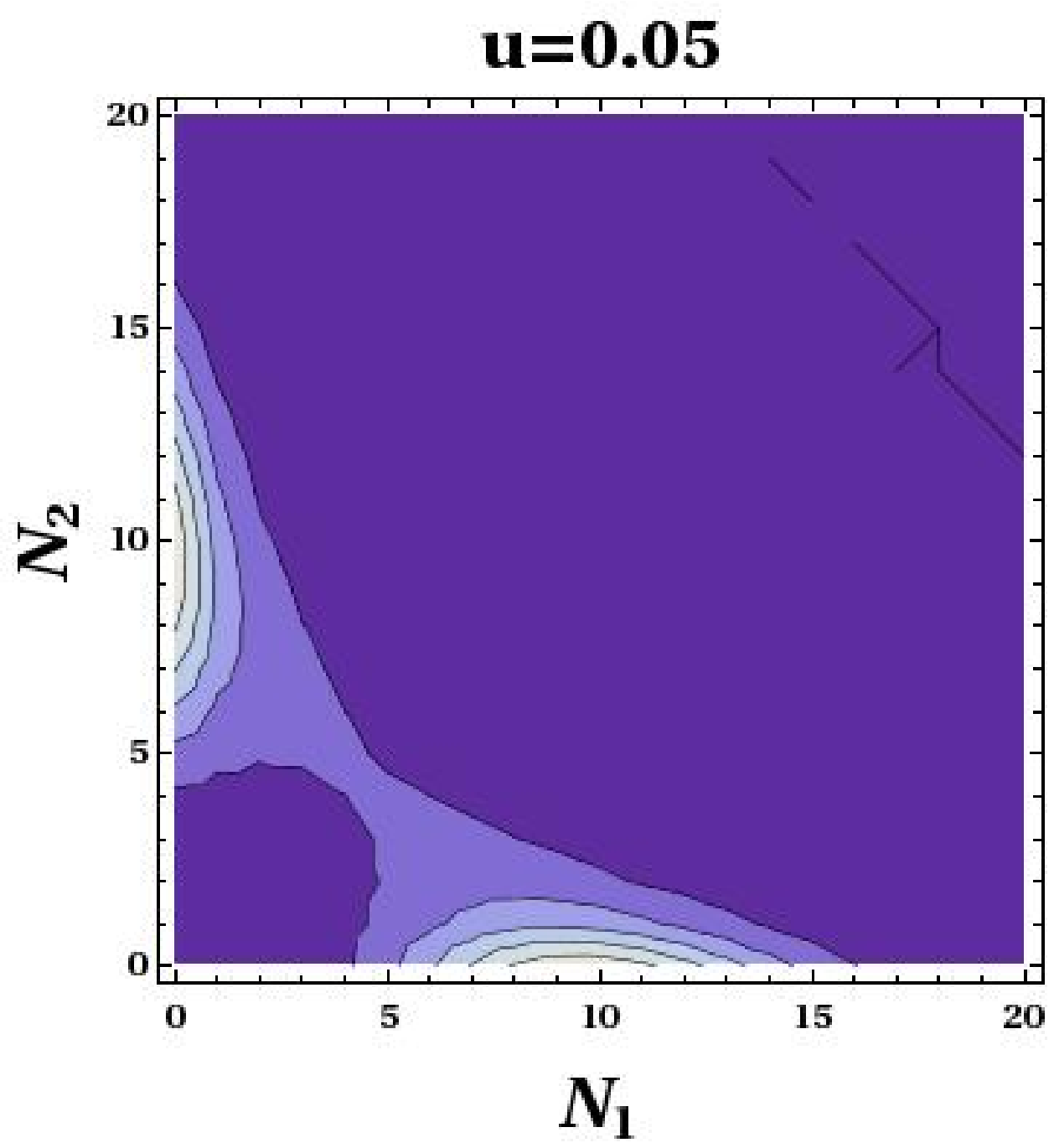}
       \textbf{a}) $u=0.05$.
    \end{center}
 \end{minipage}
 \hspace{0.02\linewidth}
 \begin{minipage}[c]{0.3\linewidth}
    \begin{center}
      
        \includegraphics[width=\linewidth]{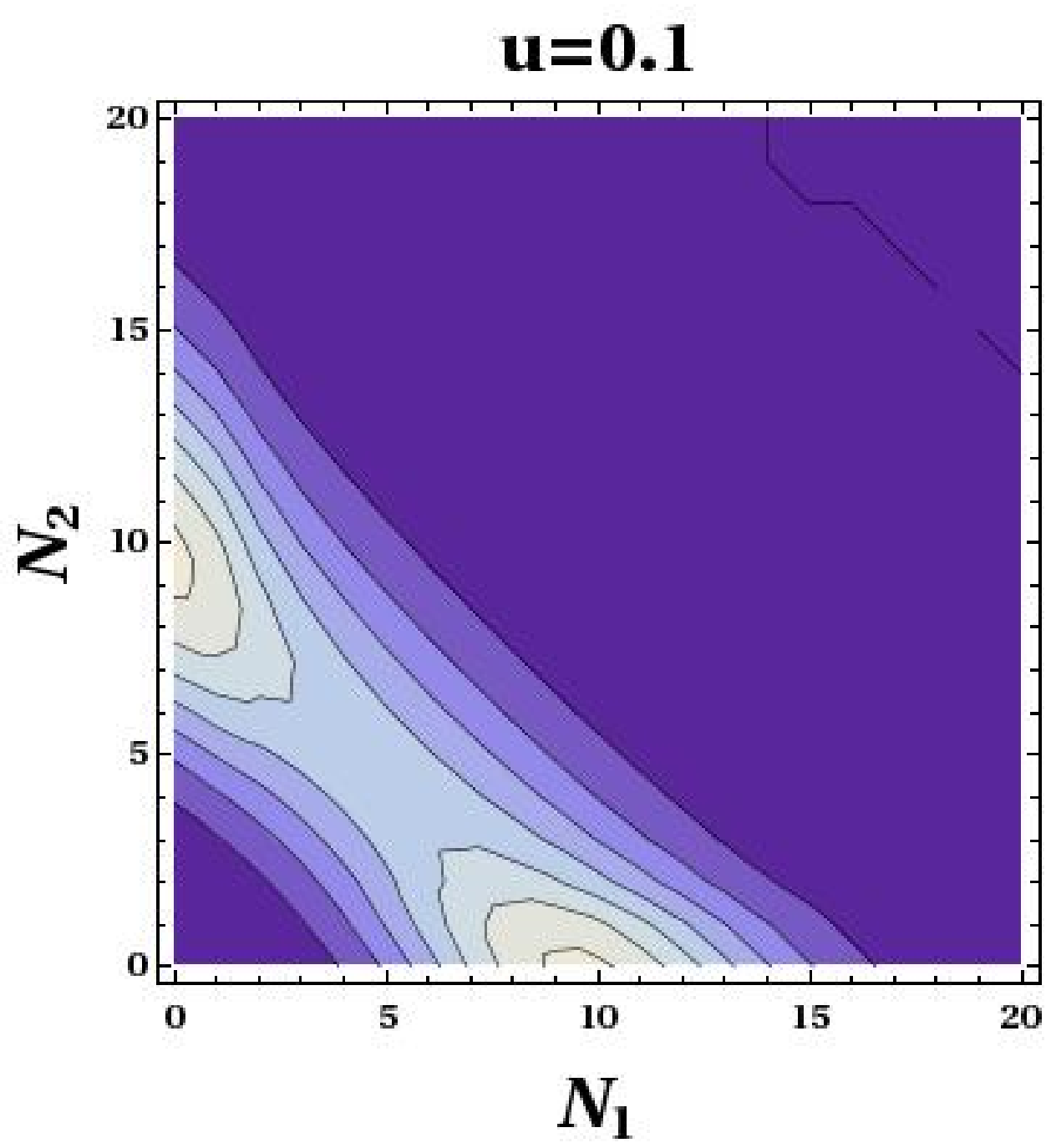}
       \textbf{b}) $u=0.1$.
    \end{center}
 \end{minipage}
 \linebreak
 \begin{minipage}[c]{0.3\linewidth}
    \begin{center}
    
        \includegraphics[width=\linewidth,]{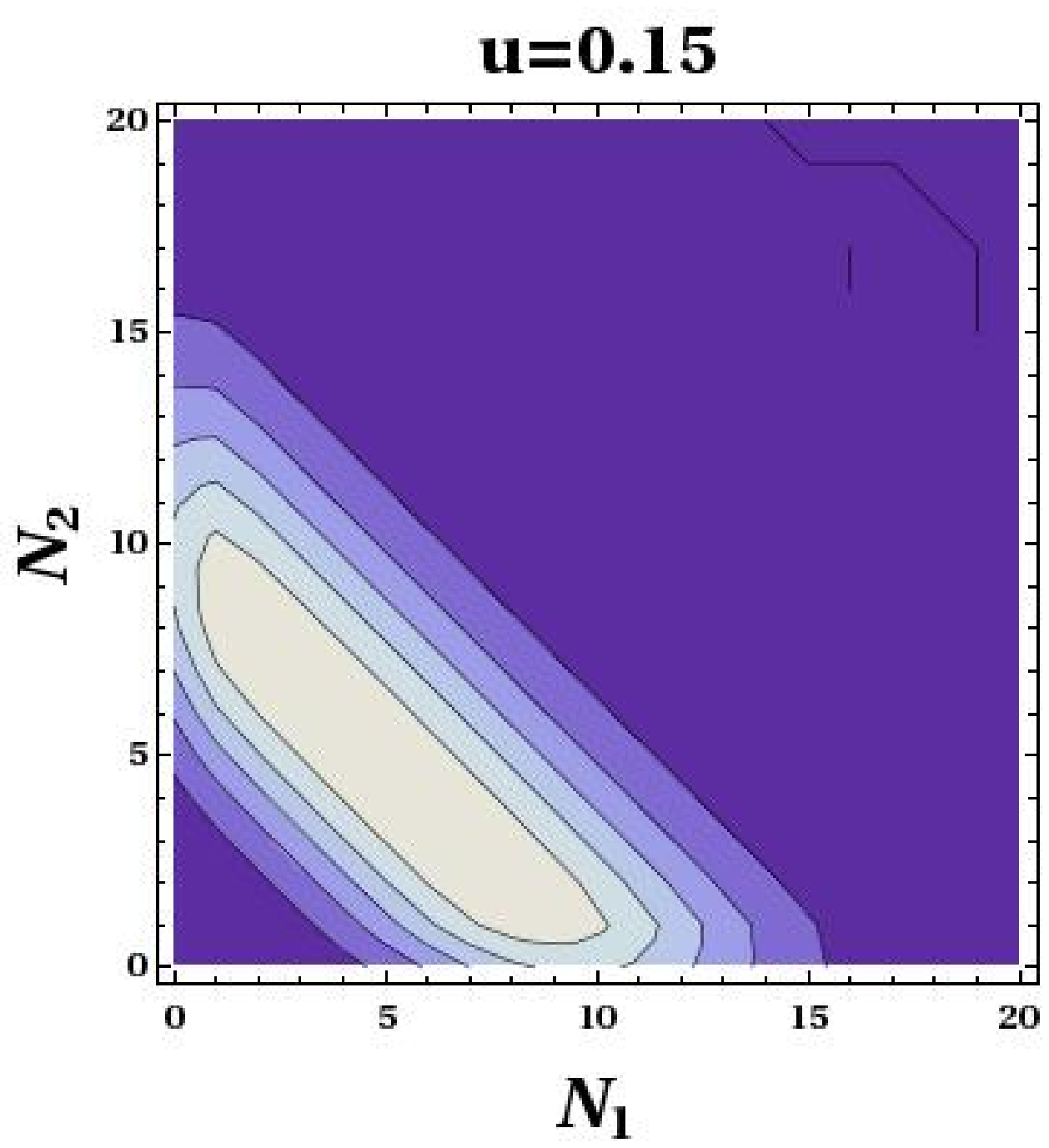}
      
    \end{center}
 \end{minipage}
\hspace{0.02\linewidth}
 \begin{minipage}[c]{0.3\linewidth}
    \begin{center}
      
        \includegraphics[width=\linewidth]{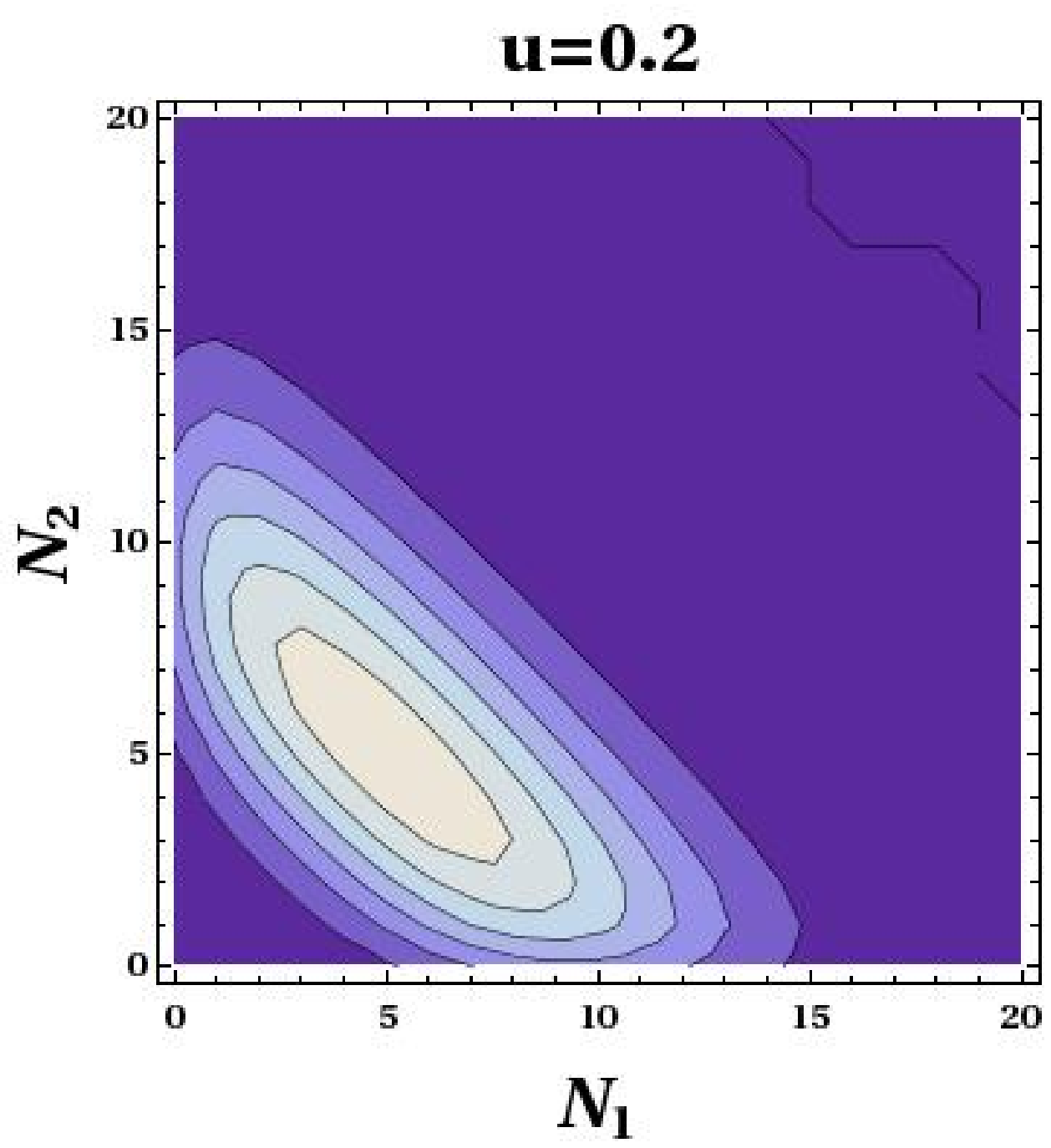}
       
    \end{center}
 \end{minipage}
\caption{
Contour plot of probability distribution $P(N_1,N_2)$
obtained from stochastic simulations.
A Transition from a two peak regime to a one peak regime 
occurs as the unbinding parameter $u$ increases. $g$,$b$ and $d$ are kept equal to the E. coli values (\ref{ecv})}
\label{fig:contour}
\end{figure}

With regard to the probability distribution $P(N_1,N_2)$, this
bistability is translated into two peaks, concentrated around the
axes, i.e, where one of the protein numbers is almost zero.  This is
illustrated in a contour plot in figure \ref{fig:contour} for
$u=0.05$.  However, this bistability depends strongly on the value of
the parameters: the bigger the value of $u$, the more irrelevant is
the switch state for the dynamics of the protein, and the less
important is the bistability we have described.  For example in figure
\ref{fig:contour}, when $g$,$b$ and $d$ are kept constant and $u$ is
increased, the peaks move together and eventually merge at some value
of $u \simeq 0.15$.  Thus there is an apparent transition from a
distribution with two symmetry-related peaks to a distribution with
one symmetric peak.  We wish to study the nature of this transition
i.e. is there an underlying phase transition at a finite value of $u$
where the system changes from symmetry-broken behaviour to symmetric
behaviour, or is the transition simply due to two peaks coming closer
together and no longer being resolved?  The latter would correspond to
a `geometrical transition' in the form of $P(N_1,N_2)$ but would not
correspond to any underlying phase transition.

\begin{figure}
\centering
\caption{
Evolution of the two  contributions to $P_1(N_1,N_2)$ defined in (\ref{rArB})---$r_A$ for $N_1>N_2$ and $r_B$ for  $N_1<N_2$---as $u$ increases.}
\label{fig:rarb}
\includegraphics[scale=0.5]{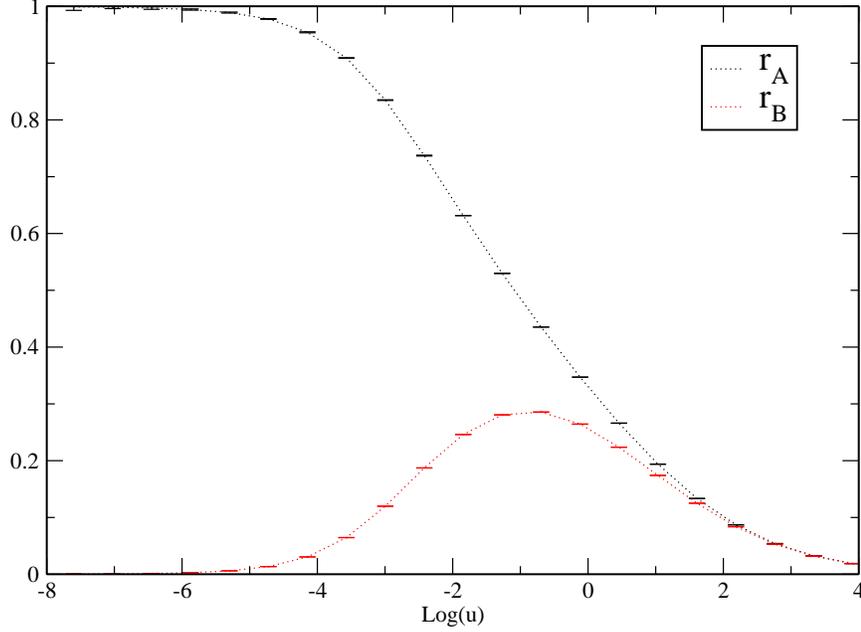}
\end{figure}

Although the whole probability distribution is always symmetric,
distributions $P_1$ and $P_2$ will be a priori asymmetric, since they
describe the probability of the number of proteins when a protein $1$
or $2$ are bound, which are not symmetric situations. Let us  now
define $r_A$ and $r_B$ as the probabilty masses of $P_1$ on either  sides of the
diagonal $N_1=N_2$ :
\begin{equation}
\eqalign{
  r_A  =\sum_{N_1>N_2}P_1(N_1,N_2)+\frac12 \sum_{N_1=N_2}P_1(N_1,N_2) \cr
  r_B  =\sum_{N_1<N_2}P_1(N_1,N_2)+\frac12 \sum_{N_1=N_2}P_1(N_1,N_2)\;.
}
\label{rArB}
\end{equation}
We can now study how $r_A$, $r_B$ change with $u$, and whether there is a clear
transition between the situation in which they are different, and the
one in which they are equal to each other (if there is any). Figure
\ref{fig:rarb} shows that these two quantities approach each other in
a continuous way, and they are equal only when $u\rightarrow \infty$, that is, when 
the only possible state of the switch is $S=0$ and it has no longer any effect 
on the protein dynamics. Therefore, the probability distributions $P_1, P_2$ and hence $r_A,
r_B$ tend to zero, all the probability being concentrated in the distribution $P_0(N_1,N_2)$, 
which is always symmetric. 

\begin{figure}[htb]
 \centering
 \begin{minipage}[c]{0.3\linewidth}
    \begin{center}
        \includegraphics[width=\linewidth]{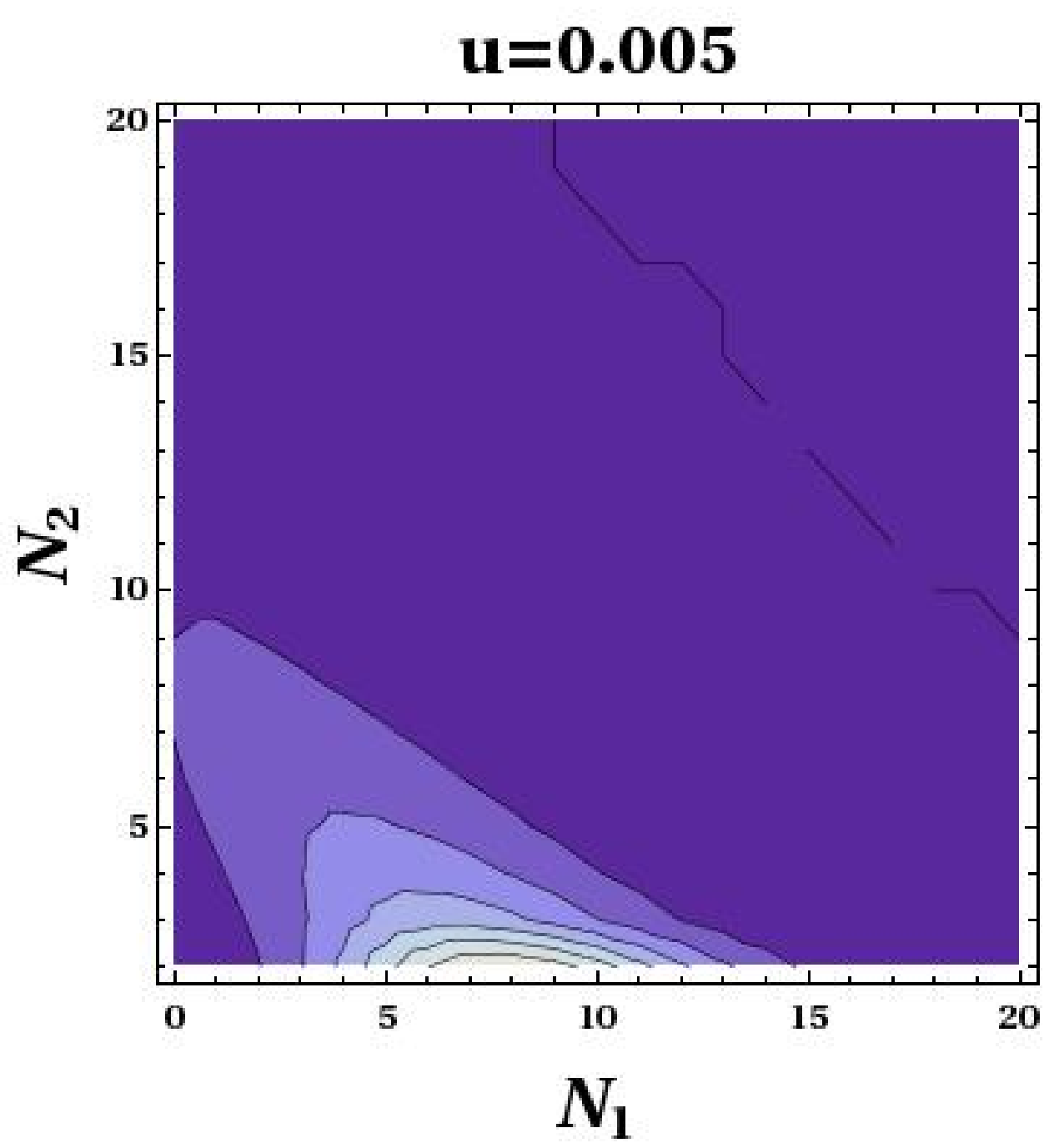}
       \textbf{a}) $u=0.005$.
    \end{center}
 \end{minipage}
 \hspace{0.02\linewidth}
 \begin{minipage}[c]{0.3\linewidth}
    \begin{center}
      
        \includegraphics[width=\linewidth]{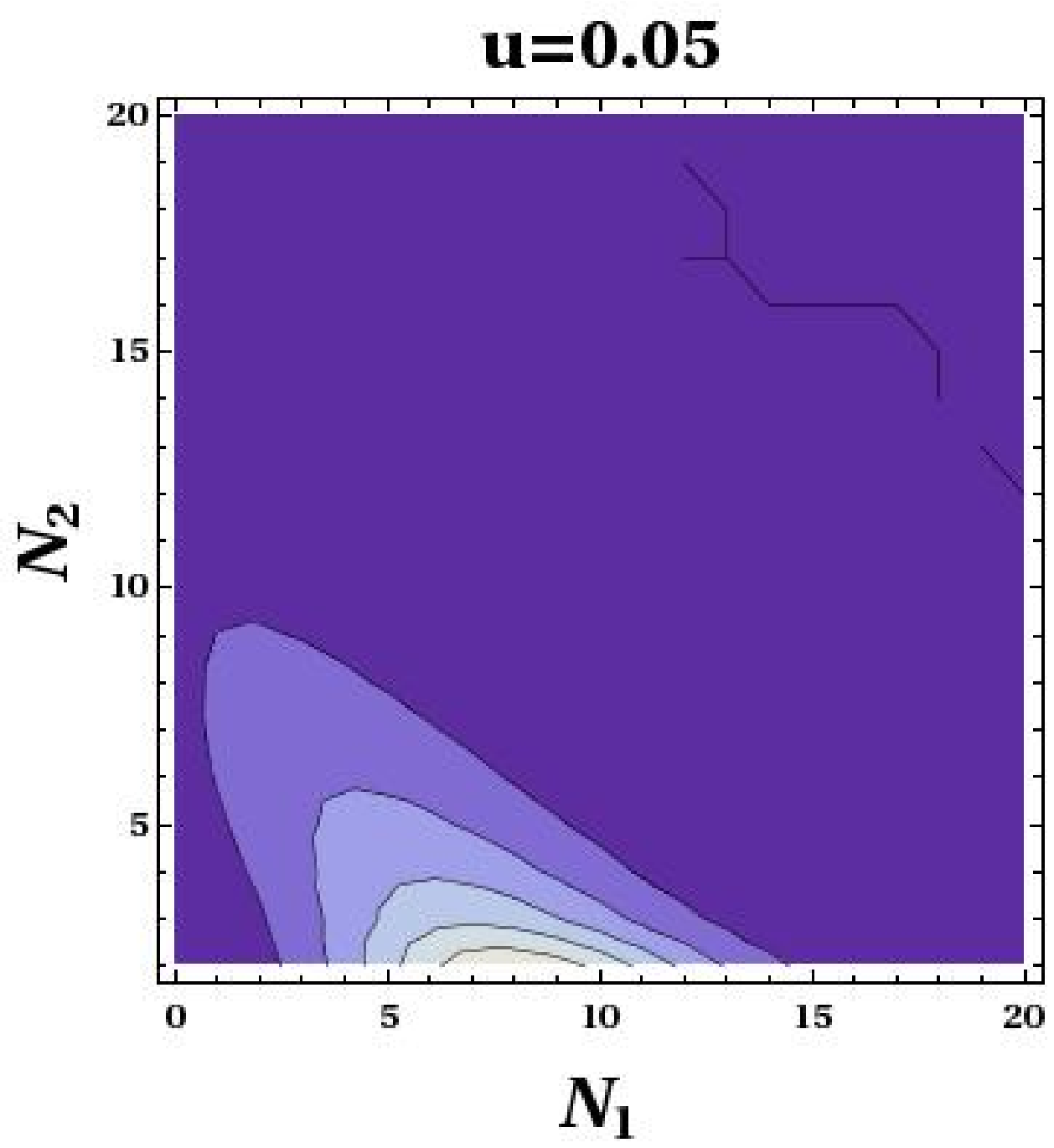}
       \textbf{b}) $u=0.05$.
    \end{center}
 \end{minipage}
 \linebreak
 \begin{minipage}[c]{0.3\linewidth}
    \begin{center}
    
        \includegraphics[width=\linewidth]{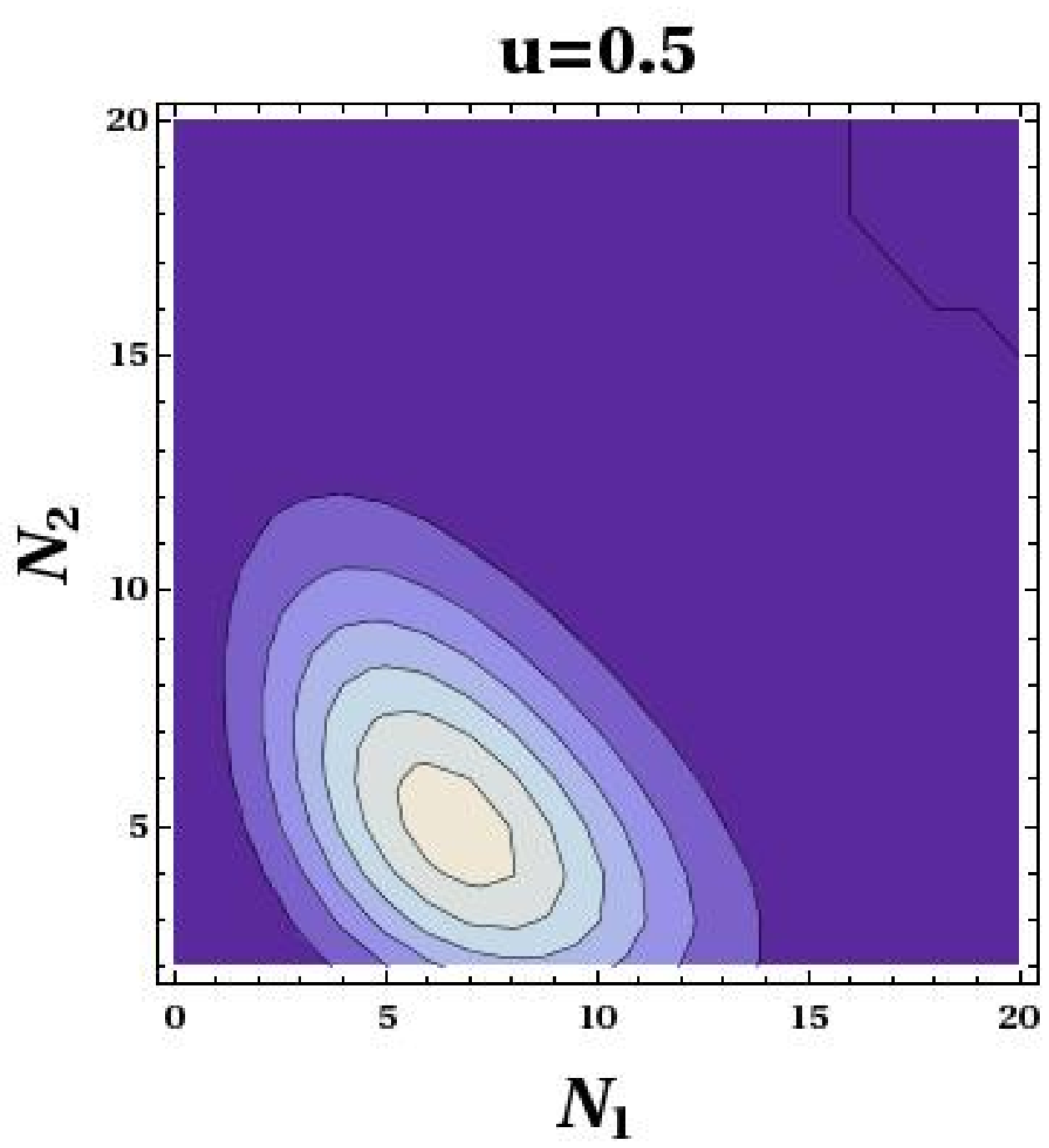}
      
    \end{center}
 \end{minipage}
\hspace{0.02\linewidth}
 \begin{minipage}[c]{0.3\linewidth}
    \begin{center}
      
        \includegraphics[width=\linewidth]{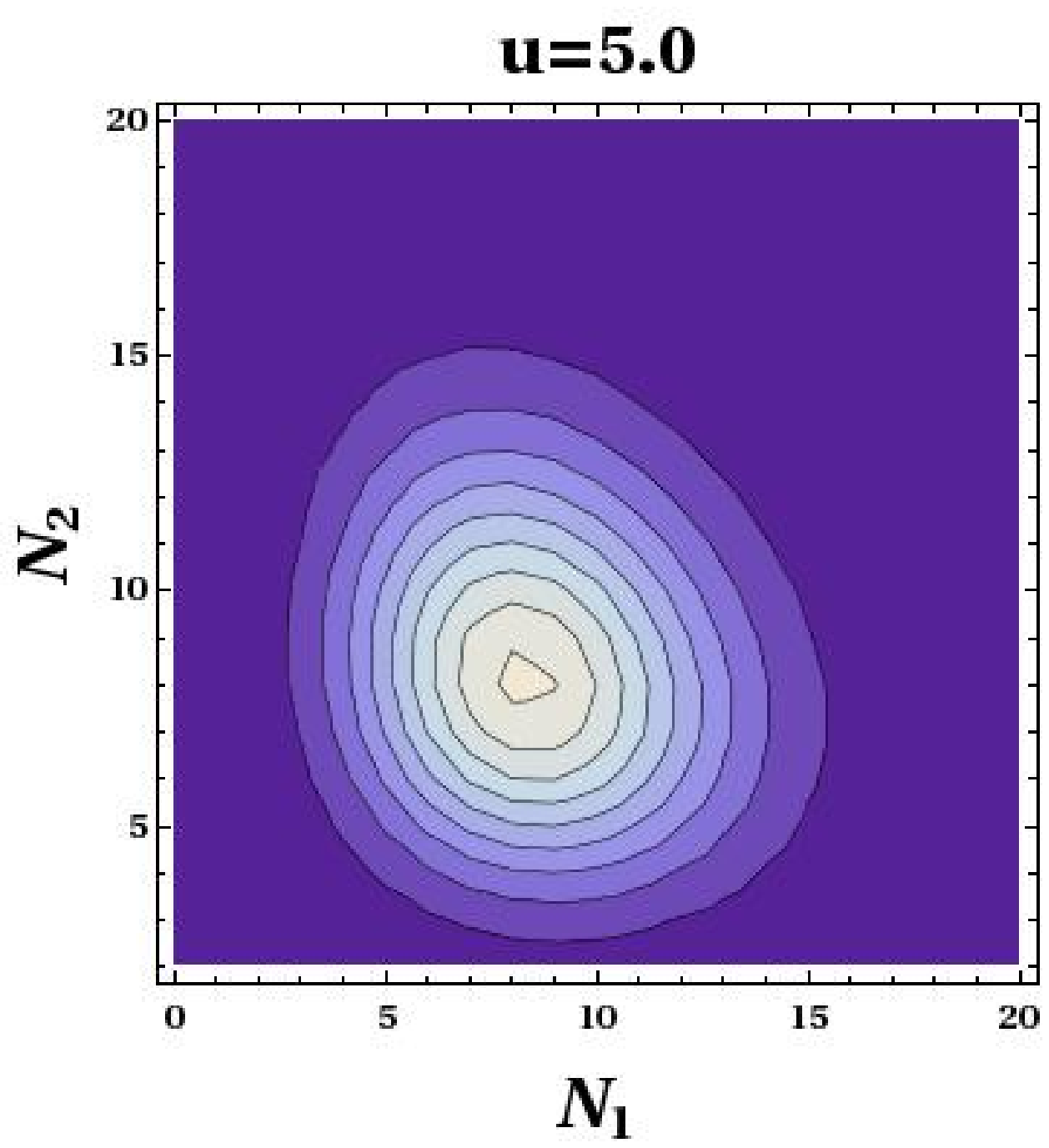}
       
    \end{center}
 \end{minipage}
\caption{Contour plots for the probability distribution $P_1(N_1,N_2)$, for different values of $u$. As $u$ increases, the peak of the distribution moves towards the
diagonal $N_1=N_2$, but the distribution is not completely symmetric as long as $u$ is finite. Also, the probability mass of the distribution decays as $u$ 
increases, and is transferred to the symmetrical distribution $P_0(N_1,N_2)$.}
\label{fig:p1dis}
\end{figure}

Despite the fact that a change in the shape of the distribution is
observed (figure \ref{fig:p1dis}), there is no evidence of the typical
singularities that appear in a phase transition.  Instead $P_1$
appears to deform continuously into a symmetric distribution when $u
\to \infty$.  We conclude that (at least for these parameter values)
$P_1(N_1,N_2)$ remains asymmetric for $u<\infty$, and as a
consequence so does $P_2$, and there is no transition to a symmetric
state in this marginal distribution. As figure \ref{fig:p1dis} shows,
$P_1$ has only one peak for different values of $u$, and even if it
becomes smaller as $u$ increases, it does not become symmetric at any
point.

We deduce that, even though $P(N_1,N_2)$ appears to become symmetric
at some finite value of $u$ (see figure \ref{fig:contour}), there is no true
phase transition between symmetric and asymmetric regimes, since $P_1$
and $P_2$ always remain asymmetric.  Thus, the bistability of the
switch is always present, with the asymmetric distributions $P_1,P_2$ 
decreasing as the switch state becomes less important,
that is, as $u$ increases.

\subsection{Solution in limit $u$, $b \to 0$}
\label{sec:ub0}
We now consider  the system
in the  limit $u$, $b \to 0$ with the binding constant
\begin{equation}
k= \frac{b}{u}
\label{bc}
\end{equation}
held fixed.
In this limit the system will equilibrate between binding/unbinding
events. When the switch is in state 1 the number of type 2 
proteins decays to zero
and the number of type 1 is given by Poissonian statistics
for the generation/degeneration processes:
\begin{equation}
P_1(N_1,N_2)
= r_1 \frac{1}{N_1!}\left(\frac{g}{d}\right)^{N_1} e^{-g/d} \delta_{N_2,0}
\end{equation}
where $r_1$ is the probability of the switch being in state 1 ($r_1=\sum_{N_1,N_2=0}^{\infty}P_1(N_1,N_2)$).
Similarly, in state 2
the number of type 2 is given by Poissonian statistics
\begin{equation}
P_2(N_1,N_2)
= r_2 \frac{1}{N_2!}\left(\frac{g}{d}\right)^{N_2} e^{-g/d} \delta_{N_1,0}
\end{equation}
and in state 0
both $N_1$ and $N_2$ are  given by Poissonian statistics
\begin{equation}
P_0(N_1,N_2)
= r_0
 \frac{1}{N_1!N_2!}\left(\frac{g}{d}\right)^{N_1+N_2} e^{-2g/d}\;.
\end{equation}
In order to fix the probabilities $r_0$,$r_1$,$r_2$
we consider the  master equation for $r_1$ which can be obtained 
from summing (\ref{master1}) over  the variables $N_1,N_2$:
\begin{equation}
\dot{r_1}
= -u r_1 + b \overline{N_1^0}
\label{r1dot}
\end{equation} 
where the quantity $\overline{N_1^0}$ 
is the mean value of $N_1$ given the system is in switch state $S=0$, multiplied
by the probability of being in switch state $S=0$, i.e, 
$\overline{N_1^0}=\sum_{N_1,N_2=0}^{\infty}N_1P_1(N_1,N_2)$.
In the stationary state 
we have  $\overline{N_1^0} = r_0 g/d$
and  (\ref{r1dot})
becomes
\begin{equation}
0 = -u r_1 + \frac{bg}{d} r_0\;.
\end{equation} 
A similar equation holds for $r_2$
and  the normalisation of probability yields
\begin{equation}
r_0 = \frac{1}{1+ 2kg/d}\qquad
r_1=r_2 =
\frac{kg/d}{1+ 2kg/d}\;.
\end{equation} 
Thus we see that  the system has three long-lived states:
when $S=1$ the system is dominated by type 1 proteins;
when $S=2$ the system is dominated by type 2 proteins, 
and when $S=0$ the system is in a symmetric state.

When $k\to \infty$ the symmetric state has zero weight ($r_0 \to 0$)
and the system is either in the $S=1$ state or the $S=2$ state.  The
switching time between the two states is expected to diverge in this
limit therefore the system exhibits symmetry breaking.  Note that the
transition from the $S=1$ state to the $S=2$ state is through the
symmetric $S=0$ state.

\subsection{Solution in limit $u$, $b \to \infty$}
\label{sec:ubinfty}
In this limit the unbinding and binding events happen
on a faster time scale than the growth and degradation  of proteins.
Therefore the switch state becomes decoupled
from the numbers of protein and
the probabilities obey
\begin{equation}
P_S(N_1,N_2) = P(N_1,N_2) r_S(N_1,N_2)
\end{equation}
where
$P(N_1,N_2)$ is the probability of
$N_1$,$N_2$ proteins being in the system regardless of switch state.
That is, for given number of proteins $N_1$ and  $N_2$
(which include any bound protein) the switch probabilities equilibrate
and obey
\begin{equation}
u r_1(N_1,N_2) = b N_1 r_0(N_1,N_2) \qquad
u r_2(N_1,N_2) = b N_2 r_0(N_1,N_2) 
\label{req2}
\end{equation}
which may be solved to  obtain
\begin{equation}
r_0 = \left[ 1+  k(N_1+N_2)\right]^{-1}
\quad r_1= k N_1 r_0 
\quad r_2= k N_2 r_0 
\end{equation}
where the binding constant $k$ is given by (\ref{bc}).

A master equation for the evolution of
$P(N_1,N_2)$ 
on the slower timescale on which generation and degeneration events occur
may then be written down:
\begin{eqnarray}
\fl \frac{\partial P}{\partial t}(N_1,N_2) = &
g[1-r_2(N_1-1,N_2) ]P(N_1-1,N_2)
 -d[N_1-r_1(N_1,N_2)]P(N_1,N_2) \label{meubinf} \\
&-g[1-r_2(N_1,N_2)P(N_1,N_2) ]
 +d[N_1+1-r_1(N_1+1,N_2)]P(N_1+1,N_2)\nonumber\\
&+ g[1-r_1(N_1,N_2-1) ]P(N_1,N_2-1)
 -d[N_2-r_2(N_1,N_2)]P(N_1,N_2)\nonumber \\
&-g[1-r_1(N_1,N_2) ]P(N_1,N_2)
 +d[N_2+1-r_2(N_1,N_2+1)]P(N_1,N_2+1)\;. \nonumber
\end{eqnarray}
The terms with coefficient $g$ in (\ref{meubinf})
represent generation  of $N_i$ when the switch is
not in switch state i.
The  terms with coefficient $d$ in (\ref{meubinf})
represent reduction   of $N_i$  with rate $dN_i$
when the switch is not  state $i$ and reduction 
with rate $d(N_i-1)$
when the switch is in  state $i$ (this is because the bound protein does not degrade).
The stationary 
state of (\ref{meubinf}) obeys detailed balance
with respect to generation and degradation of
$N_1$ and $N_2$ individually, thus
\begin{eqnarray}
P(N_1,N_2)&=&
P(N_1-1,N_2)\frac{g}{d}\frac{[1-r_2(N_1-1,N_2) ]}{[N_1-r_1(N_1,N_2)]} \\
&=&
P(N_1-1,N_2)\frac{g}{d}\frac{[1+k(N_1-1)]}{[1+k(N_1+N_2-1)]}
\frac{1}{N_1} \frac{[1+k(N_1+N_2)]}{[1+k(N_1+N_2-1)]}\nonumber
\label{Pubinf}
\end{eqnarray}
with a similar equation holding for detailed balance in $N_2$.
Then these equations may be iterated to obtain
\begin{eqnarray}
\fl P(N_1,N_2)
 &=&
\left(\frac{g}{d}\right)^{N_1}
 \frac{(1+k(N_1+N_2))}{(1+kN_2)}
\frac{1}{N_1!}
\prod_{n_1=0}^{N_1-1}\frac{(1+k n_1)}{1+k(N_2+ n_1)}
P(0,N_2)\nonumber \\
\fl &=&
\left(\frac{g}{d}\right)^{N_1+N_2}
 \frac{(1+k(N_1+N_2))}{N_1! N_2!}
\frac{\prod_{n_1=0}^{N_1-1}(1+k n_1)
\prod_{n_2=0}^{N_2-1}(1+k n_2)}{\prod_{n=0}^{N_1+N_2-1}(1+k n)}
P(0,0)\;.
\label{Pubinf2}
\end{eqnarray}
The constant $P(0,0)$ will be determined by normalisation
of the sum of probabilities to one.

The expression (\ref{Pubinf})
is a single-peaked distribution
which is symmetric in $N_1$ and $N_2$.
To see this one can  identify the stationary points
of $P(N_1,N_2)$ through the conditions
$P(N_1+1,N_2) =P(N_1,N_2)$ and $P(N_1,N_2+1)= P(N_1,N_2)$
which yield
\begin{equation}
\eqalign{
\frac{g}{d} \frac{1}{(N_1+1)}
\frac{(1+kN_1)}{(1+k(N_1+N_2))}
\frac{(1+k(N_1+N_2+1))}{(1+k(N_1+N_2))} &=1
\cr
\frac{g}{d} \frac{1}{(N_2+1)}
\frac{(1+kN_2)}{(1+k(N_1+N_2))}
\frac{(1+k(N_1+N_2+1))}{(1+k(N_1+N_2))} &=1\;.
}
\end{equation}
The solution of these equations
is $N_1=N_2= N/2$
where,
when  $N$ is large,
\begin{equation}
kN^2 + \left[1- k\frac{g}{d}\right]N - 2 \frac{g}{d} =0\;.
\label{Nubinf}
\end{equation}
Thus, in the $u,b\to \infty$ limit
the system has reached a symmetric state
with the  probability distribution
peaked at $N_1=N_2= N/2$.



\section{Mean Field Theory}
\label{sec:mft}
In this section we develop a mean field theory in which some
correlations in the numbers of proteins are ignored.  

\subsection{Exact Moment Equations}
We start from the exact master equations (\ref{master0}--\ref{master2})
for the evolution of probabilities $P_S(N_1,N_2)$.
The zeroth moments of $N_i$ are the probabilities $r_S$
i.e.
\begin{equation}
r_S = 
\sum_{N_1=0,N_2=0}^{\infty} P_S(N_1,N_2)\qquad\mbox{for} \quad S= 0,1,2\;.
\label{ri}
\end{equation}
We now define the first moments of $\overline{N_i^S}$ of $N_i$ as follows
\begin{equation}
\overline{N_i^S} =
\sum_{N_1=0,N_2=0}^{\infty} N_i P_S(N_1,N_2)\qquad\mbox{for}\quad i=1,2\quad S= 0,1,2
\label{Nia}
\end{equation}
and the second moments
\begin{equation}
\overline{(N_iN_j)^S}=
\sum_{N_1=0,N_2=0}^{\infty} N_i N_j P_S(N_1,N_2)\qquad\mbox{for}\quad i,j=1,2\quad S= 0,1,2\;.
\label{Nija}
\end{equation}
Summing (\ref{master0}) gives
\begin{eqnarray}
\frac{\partial r_0}{\partial t} &=&
-b(\overline{N_1^0}+ \overline{N_2^0})
+u[r_1+r_2]
\label{r0t}\\
\frac{\partial \overline{N_1^0}}{\partial t} &=&gr_0
-d\overline{N_1^0}
 -b[\overline{(N_1N_1)^0}+\overline{(N_1N_2)^0}]
+ u[\overline{(N_1)^1}+\overline{(N_1)^2}+r_1]\;.
\label{N10t}
\end{eqnarray}
Note that the physical meaning of e.g.
$\overline{N_i^S}$ is
the probability of being in switch state $S$ ($r_S$) multiplied by the mean number of type $i$, {\em given} that the switch is in state $S$.

Similarly, summing (\ref{master1}) gives
\begin{eqnarray}
\frac{\partial r_1}{\partial t} &=&
+b\overline{N_1^0} -ur_1
\label{r1t}\\
\frac{\partial \overline{N_1^1}}{\partial t} &=&gr_1
-d\overline{N_1^1}
 +b\overline{(N_1(N_1-1))^0}
- u\overline{(N_1)^1}
\label{N11t}\\
\frac{\partial \overline{N_2^1}}{\partial t} &=&
-d\overline{N_2^1}
 +b\overline{(N_1N_2)^0}
- u\overline{(N_2)^1}\;.
\label{N21t}
\end{eqnarray}
We now invoke symmetry
between switch state 1 and 2 :
\begin{equation}
r_1=r_2 = (1-r_0)/2\quad
\overline{N_1^0} =\overline{N_2^0} \quad
\overline{N_1^1} =\overline{N_2^2} \quad
\overline{N_1^2} =\overline{N_2^1}\;. 
\label{symm}
\end{equation}
Then the exact steady-state versions of equations
(\ref{r0t}--\ref{N21t}) read
\begin{eqnarray}
 r_1=
\frac{b}{u}\overline{N_1^0} \qquad r_0 = 1-2 r_1
\label{rss}\\
b[\overline{(N_1N_1)^0}+\overline{(N_1N_2)^0}]
=gr_0
-d\overline{N_1^0}
+ u[
\overline{(N_1)^1}+ \overline{(N_1)^2}+r_1]
\label{N10ss}\\
b\overline{(N_1N_1)^0}
=-gr_1
+(d+u)\overline{N_1^1}
+ b\overline{(N_1)^0}
\label{N11ss}\\
b\overline{(N_1N_2)^0}
=(d+u)\overline{N_2^1}\;.
\label{N12ss}
\end{eqnarray}
Note that if we sum (\ref{N10ss}--\ref{N12ss}) we obtain the exact relation
\begin{equation}
d(  \overline{N_1^0}+\overline{N_1^1}+\overline{N_1^2}) =g(1-r_1)=g(1-r_2)
\label{Nss}
\end{equation}
which simply  gives the overall birth/death balance for $N_1$.
Also note that
(\ref{N11ss},\ref{N12ss}) give exact relations between the second moments and first moments. However to actually evaluate these moments one would have to consider equations for higher moments, leading to a hierarchy of equations.

\subsection{Mean Field Approximation}
We now make a mean-field approximation that 
expresses second moments in terms of first moments:
\begin{eqnarray}
\overline{(N_1N_1)^0}
&=& \frac{ \overline{N_1^0}\ \overline{N_1^0}}{r_0} 
+\overline{N_1^0}\label{N1N1mf}\\
\overline{(N_1N_2)^0}
&=& \frac{ \overline{N_1^0}\ \overline{N_2^0}}{r_0}=\frac{ (\overline{N_1^0})^2}{r_0}\;.
\label{N1N2mf}
\end{eqnarray}
Note that a symmetry condition from  (\ref{symm}) has been explicitly used in the last equation of (\ref{N1N2mf}).
The first  relation (\ref{N1N1mf}) comes from the assumption that
$N_1$  has a Poisson distribution when the switch is in the 0 state. This means that
the second moment is equal to the square of the mean plus the mean itself. However, there is an important factor $r_0$, which comes from the fact that
$\overline{(N_1N_1)^0}/r_0$ is the mean square value of $N_1$ {\em given}
that the switch is in the 0 state
and $\overline{N_1^0}/r_0$ is the
mean  value of $N_1$ {\em given}
that the switch is in the 0 state.
The Poisson approximation is in fact exact in the limit where
$u$,$b$ tend to zero (see section \ref{sec:ub0}).

The second relation (\ref{N1N2mf}) is a simple factorization scheme
which ignores correlations between the values of $N_1$ and $N_2$ when
the switch state is 0.

Using this approximation scheme (\ref{N11ss}) becomes 
\begin{equation}
\overline{N_1^1}
=\frac{b}{d+u} \frac{(\overline{N_1^0})^2}{r_0} + \frac{gr_1}{d+u} 
=\frac{br_0}{d+u}\left[
 \left(\frac{\overline{N_1^0}}{r_0}\right)^2 + \frac{g}{u} \left(\frac{\overline{N_1^0}}{r_0}\right) \right]
\label{N11mf}
\end{equation}
and  (\ref{N12ss}) becomes
\begin{equation}
\overline{N_1^2}
= \frac{br_0}{d+u} \left(\frac{\overline{N_1^0}}{r_0}\right)^2\;.
\label{N12mf}
\end{equation}
Using expressions (\ref{N11mf}) and (\ref{N12mf})  in
(\ref{N10ss}), combined with the previous approximations (\ref{N1N1mf}) and (\ref{N1N2mf})
yields the following quadratic equation for $\overline{N_1^0}/r_0$:
\begin{equation}
\frac{2bd}{d+u}\left(\frac{\overline{N_1^0}}{r_0}\right)^2+\left[d-\frac{bg}{d+u}\right] \left(\frac{\overline{N_1^0}}{r_0}\right) -g=0\;.
\label{N10mf}
\end{equation}
One must take the positive root of this quadratic which yields
\begin{equation}
\frac{\overline{N_1^0}}{r_0}
= \frac{1}{4bd}\left[
bg -d(d+u) + \left( (d(d+u)-bg)^2 + 8bdg(d+u)\right)^{1/2}\right]\;.
\end{equation}
Then using (\ref{rss})
one obtains
\begin{eqnarray}
r_0 &=& \left[ 1 + \frac{2b}{u}\frac{N_1^0}{r_0}\right]^{-1}\\
\overline{N_1^0} &=&  \left(\frac{\overline{N_1^0}}{r_0}\right)\left[ 1 + \frac{2b}{u}\frac{\overline{N_1^0}}{r_0}\right]^{-1}\;.
\end{eqnarray}

One may check the  limits of section 2 from the quadratic
(\ref{N10mf}).
In the limit $b$,$u$ $\to 0$ with $k=b/u$ 
one obtains
$\overline{N_1^0}/r_0 = g/d$
and  $r_0 = \left[1 + \frac{2kg}{d}\right]^{-1}$
in agreement with
Section  2.3 where it is shown that  $N_1$ follows a Poisson distribution with mean $g/d$ when the switch is in state $S=1$ or $S= 0$.

In the limit $b$,$u$ $\to \infty$ with $k=b/u$  fixed
the quadratic (\ref{N10mf}) reduces 
to  
\begin{equation}
\left(\frac{\overline{N_1^0}}{r_0}\right)^2 2kd + 
\frac{\overline{N_1^0}}{r_0} (d- kg) -g =0\;.
\end{equation}
This quadratic for
$\frac{\overline{N_1^0}}{r_0}$ is the same
as the quadratic (\ref{Nubinf}) for the value of  $N=2N_1$ 
that maximises $P(N_1,N_2)$ in the exact solution
of section \ref{sec:ubinfty}.


\subsection{Comparison to simulation results}
The mean field theory we have developed
can be compared with the simulations by studying the zeroth and first order moments of 
the probabilty distributions, i.e., $r_0,\overline{N_1^0},\overline{N_1^1},\overline{N_1^2}$. (The probabilities 
$r_1$ and $r_2$ may  automatically be obtained from $r_0$,
 since $r_1=r_2=\frac{1-r_0}{2}$.)

Different values of the mentioned quantities are given in table $1$, where the parameters of the models have different values. 
The reference for this table is the set of E. coli values, and only the parameters that are changed are written.
 For the simulations performed, $r_0$ is always in good agreement with the mean field theory approximation, and so are $r_1$ and $r_2$. Also, 
$\overline{N_1^0}$ is in quite good agreement, too.
\begin{center}
\begin{table}
    \begin{tabular}{|l || l | l || l | l ||} 
    \hline
      & E. coli & MFT & u=0.05  & MFT  \\
\hline		
$r_0$       & $(4.9186 \pm 0.0007)\cdotp 10^{-3}$  & $4.92705\cdotp 10^{-3}$ &  $(4.5263\pm 0.0008)\cdotp 10^{-2}$ & $4.54635\cdotp 10^{-2}$\\
$\overline{N_1^0}$     & $(2.489\pm 0.005)\cdotp 10^{-2}$ & $2.48768\cdotp 10^{-2}$ & $0.23862\pm 0.00014$ & $0.238634$ \\
$\overline{N_1^1}$     & $4.942\pm 0.008$ & $3.74372$          & $4.113\pm 0.003$ & $2.71128$    \\
$\overline{N_1^2}$     & $(5.908 \pm 0.007)\cdotp 10^{-2}$ & $1.25604$    & $0.8734\pm 0.0005$ & $2.2774$  \\
    \hline \hline
& $u=50,b=1000$ & MFT  & $u=5\cdotp 10^{-8}, b=10^{-6}$  & MFT \\
\hline
$r_0$       & $(4.941 \pm 0.004)\cdotp 10^{-3}$ &  $4.95073\cdotp 10^{-3}$  & $(2.48 \pm 0.03)\cdotp 10^{-3}$  & $2.49872\cdotp 10^{-3}$\\
$\overline{N_1^0}$     & $(2.55 \pm 0.06)\cdotp 10^{-2}$ &  $2.48762\cdotp 10^{-2}$  & $(2.477\pm 0.024)\cdotp 10^{-2}$  & $2.49375\cdotp 10^{-2}$\\
$\overline{N_1^1}$     & $9.89\pm 0.11$ &  $2.50019$        & $4.92\pm 0.05$    & $4.98751$\\
$\overline{N_1^2}$      & $0.227\pm 0.013$ &  $2.49969$     & $(5.012\pm 0.007)\cdotp 10^{-5}$  & $4.97754\cdotp 10^{-5}$\\
    \hline 
    \end{tabular}
\caption{Results from different quantities from simulations and mean field theory approach. The values of $r_0$ and $\overline{N_1^0}$ predicted by 
the mean field theory are always in good agreement with the simulations. $\overline{N_1^1}$ and $\overline{N_1^2}$ predictions are quite far from the simulation results, 
except on the limit $u,b\rightarrow 0$, where our approximation is exact.}
\end{table}
\end{center}
$\overline{N_1^1}$ and $\overline{N_1^2}$ are different in the mean field theory, which is an improvement over simpler approximations where they have the same 
value. However, the values of
$\overline{N_1^1}$ and $\overline{N_1^2}$ are rather different from  the simulations, and this comes from ignoring higher correlations of the numbers of proteins and the state
of the switch. Only when $u,b\rightarrow 0$ are the values  in close agreement
with the simulation values as expected in this limit
where the mean field theory is exact.

Mean field theory can, in principle,  be improved by considering higher order moments
and correlations. However, 
the algebra soon gets quite complicated.


\section{Exact perturbative solution}
In this section we develop a perturbative approach
that allows the steady state probabilities  $P_S(N_1,N_2)$ to be computed 
as a power series in the unbinding rate $u$.

\subsection{Formal solution}
We begin
by considering the formal solution of the master equation system 
(\ref{master0}--\ref{master2}).
To transform the system of equations into a system of partial differential equations, we take the generating function of the different probability distributions:
\begin{equation}
 K_{S}(z_1,z_2)=\sum^{\infty}_{N_1=0}\sum^{\infty}_{N_2=0}z_1^{N_1}z_2^{N_2}P_{S}(N_1,N_2)
\label{Gdef}
\end{equation}
where $S=0,1,2$. In this way we obtain a system of linear partial differential equations with non-constant coefficents:
\begin{eqnarray}
 g(z_1+z_2-2)K_0+ \left[d-(d+b)z_1\right]
\frac{\partial K_0}{\partial z_1}
+\left[d-(d+b)z_2\right]\frac{\partial K_0}{\partial z_2}\nonumber&& \\
\hspace{7cm}+uz_1K_1+uz_2K_2 & =&0\label{K0ss}\\
 \left[g(z_1-1)-u\right]K_1+d(1-z_1)\frac{\partial K_1}{\partial z_1}
+d(1-z_2)\frac{\partial K_1}{\partial z_2}
+b\frac{\partial K_0}{\partial z_1} &=& 0\label{K1ss}\\
\left[g(z_2-1)-u\right]K_2+d(1-z_1)\frac{\partial K_2}{\partial z_1}+d(1-z_2)
\frac{\partial K_2}{\partial z_2}+b\frac{\partial K_0}{\partial z_2} &=&0
\label{K2ss}
\end{eqnarray}
where the right-hand side terms have been set to $0$, since the stationary probabilities $P_S(N_1,N_2)$ are the main quantities to be determined in this paper.

The second and the third equation of the system work in a completely
analogous way, because of the symmetry of species $1$ and $2$, so it
will be enough to deal with the first two equations. 
We also note the symmetries
 $K_2(z_1,z_2)=K_1(z_2,z_1)$ 
and  $K_0(z_1,z_2)=K_0(z_2,z_1)$.

In appendix $A$ we give a formal solution to the system
(\ref{K0ss}--\ref{K2ss}). However, in practice it is not clear how to
actually compute e.g. probability distributions from this solution.
In order to do this we develop instead a perturbative approach.

\subsection{Perturbative approach}
In this section we develop a perturbative approach to the problem of
finding the exact stationary state.  To do so we require a suitable
small parameter of the model which we choose to be $u$, the
unbinding parameter.  In the $u \to 0$, the exact solution is simple:
if, for example, one protein of type $1$ is bound, the proteins of
this kind will obey the usual Poisson distribution regulated by death
and birth terms, while the number of proteins of type $2$ will just
decay to $0$. This limit is the starting
point for  a perturbative solution, wherein
the probability distribution will be expanded in a
power series of $u$:
\begin{equation}
P_S = \sum_{n=0}^{\infty}  u^n P_S^{(n)}\;.
\label{Ppert}
\end{equation}
Owing to the symmetry of the system $P_2(N_1,N_2)=P_1(N_2,N_1)$
we need  only  consider $P_0$ and $P_1$.

Writing out the expansion explicitly we have
\begin{equation}
\eqalign{
  P_1 &= P_1^{(0)}+ uP_1^{(1)} \dots\\
  P_0 &=uP_0^{(1)}+ u^2 P_0^{(2)} \dots
}
\end{equation}
Note that the constant term 
$P_0^{(0)}=0$ since $P_0 =0$ in the limit of no unbinding.

This approach also makes sense when the typical E. coli values for the parameters are considered \cite{BB08}:
\begin{equation}
 g=0.05 \quad d=0.005 \quad b=0.1 \quad u=0.005 
\end{equation}
$u$ is, along with $d$, the smallest of the parameters. An expansion in $1/b$ 
could  also be developed, but we find the expansion in $u$ more convenient.

\subsection{Zeroth order}
In the zeroth order of the $u$ expansion of the stationary master 
equation (\ref{master2}) (with l.h.s set to zero) we find
\begin{eqnarray}
\fl 0&=&g[P_1^0(N_1-1,N_2)-P_1^0(N_1,N_2)]+d[(N_1+1)P_1^0(N_1+1,N_2)+(N_2+1)P_1^0(N_1,N_2+1)\nonumber\\
\fl  &&-(N_1+N_2)P_1^0(N_1,N_2)]\;.
\end{eqnarray}
We define the generating function
\begin{equation}
K_1^{(0)}(z_1,z_2) =  
\sum^{\infty}_{N_1=0}\sum^{\infty}_{N_2=0} z_1^{N_1}z_2^{N_2}P_{1}^{(0)}(N_1,N_2)
\end{equation}
which obeys
\begin{equation}
 0=g(z_1-1)K^{(0)}_1+
d(1-z_1)\frac{\partial K^{(0)}_1}{\partial z_1} +
d(1-z_2)\frac{\partial K^{(0)}_1}{\partial z_2} 
\end{equation}
the solution of which is independent of $z_2$:
\begin{equation}
 K^{(0)}_1 = c_1\exp{\frac{g}{d}z_1}
\end{equation}
where $c_1$ is a constant to be determined. Analogously
$K_2^{(0)}(z)=c_2\exp{\frac{g}{d}z_2}$. 
Applying the normalization condition $K^{(0)}_1(1)+K^{(0)}_2(1)=1$ and the symmetry consideration $c_1=c_2$ leads to 
\begin{equation}
K_1^{(0)}=\frac12\exp{-\frac{g}{d}}\ \exp{\frac{g}{d}z_1}\;. 
\end{equation}
Expanding as a power series in $z_1,z_2$ yields
\begin{equation}
P_1^{(0)}(N_1,N_2)=\frac12\frac{\exp{-\frac{g}{d}}}{N_1!}\left(\frac{g}{d}\right)^{N_1}\delta_{N_2,0}\;.
\label{P10}
\end{equation}
This is a Poisson distribution for $N_1$ with mean $g/d$, with $N_2$ fixed to be zero.
The normalisation factor $1/2$ is so that
$P^{(0)}(N_1,N_2)= P_1^{(0)}(N_1,N_2) +
P_2^{(0)}(N_1,N_2)$  is normalised to unity.

\subsection{General Formulation}
\label{sec:genform}
We  substitute the expansion (\ref{Ppert}) into the stationary
master  system  (\ref{master0}--\ref{master2}) with time derivatives set
equal to zero. Arranging orders of $u$ 
the equations may be written as
\begin{equation}
\mathcal{L}_S P_S^{(n)}(N_1,N_2) = -f_S^{(n)}(N_1,N_2)
\label{LPf}
\end{equation}
for $S=0,1$ where the action of the linear operators
$\mathcal{L}_S$ is
\begin{eqnarray}
\fl \mathcal{L}_0 P_0^{(n)}(N_1,N_2) =&g[P_0^{(n)}(N_1-1,N_2)+P_0^{(n)}(N_1,N_2-1)-2P_0^{(n)}(N_1,N_2)]\nonumber \\
&+d\left[(N_1+1)P_0^{(n)}(N_1+1,N_2)+(N_2+1)P_0^{(n)}(N_1,N_2+1)\right. \nonumber \\
&\left. -(N_1+N_2)P_0^{(n)}(N_1,N_2)\right]-b(N_1+N_2)P_0^{(n)}(N_1,N_2)\label{LP0}\\
\nonumber \\ 
\fl \mathcal{L}_1 P_1^{(n)}(N_1,N_2)  =&g[P_1^{(n)}(N_1-1,N_2)-P_1^{(n)}(N_1,N_2)]
\nonumber \\
&+d\left[(N_1+1)P_1^{(n)}(N_1+1,N_2)+(N_2+1)P_1^{(n)}(N_1,N_2+1)\right. \nonumber \\
& \left.-(N_1+N_2)P_1^{(n)}(N_1,N_2)\right]
\label{LP1}
\end{eqnarray}
and the inhomogenous terms are 
\begin{eqnarray}
f_0^{(n)}(N_1,N_2)&=& P_1^{(n-1)}(N_1-1,N_2)+P_2^{(n-1)}(N_1,N_2-1) \label{f0}\\
f_1^{(n)}(N_1,N_2)&=&-P_1^{(n-1)}+b(N_1+1)P_0^{(n)}(N_1+1,N_2) \label{f1}\;.
\end{eqnarray}
As noted above, we  can first determine
the zeroth order $P_1^{(0)}(N_1,N_2)$ and $P_2^{(0)}(N_1,N_2)
=P_1^{(0)}(N_2,N_1)$ as functions  of
the parameters of the model. Then,
owing to the form of the equation (\ref{f0}),  $f_0^{(1)}$ is determined.
This allows us to  solve for $P_0^{(1)}$ which in turn determines
$f_1^{(1)}$ and allows us to solve for $P_1^{(1)}$.
Continuing in this fashion
the rest of the probabilities will
be found following the  structure:
\begin{equation}
P_1^{(0)}\rightarrow P_0^{(1)} \rightarrow P_1^{(1)} \rightarrow P_0^{(2)}
\rightarrow P_1^{(2)} \cdots\end{equation} 
In general, $P_0^{(n)}$ will be
found before $P_1^{(n)}$.

Having laid out the general perturbation scheme 
and established the zeroth order ($n=0$) solution,
 we now outline how
equations (\ref{LPf}) can be solved.  We note that for each switch
state $S$ the same linear operator $\mathcal{L}_S$ appears at all orders
$n$.
This means that the homogeneous parts of the equations are independent of the order (with only the inhomogeneous term on the right hand side varying between
orders)
and that they only have to be solved once.

\subsection{Green function for $\mathcal{L}_0$}

Let us define a Green Function $Q_0(N_1,N_2|N^0_1,N^0_2)$ for
the  operator $\mathcal{L}_0$ through
\begin{equation}
\mathcal{L}_0 Q_0(N_1,N_2|N^0_1,N^0_2)  = -\delta_{N_1,N_1^0}\delta_{N_2,N_2^0}
\label{Q}
\end{equation}
so that the solution of
(\ref{LPf}) may be written
\begin{equation}
P_{0}^{(n)}(N_1,N_2)=
\sum_{N_1^0,N_2^0}f_0^{(n)}(N_1^0,N_2^0)Q_0(N_1,N_2|N_1^0,N_2^0)\;.
\label{greensol}\end{equation}

We define a generating function
\begin{equation}
K_0(z_1,z_2|N_1^0,N_2^0) = 
\sum^{\infty}_{N_1=0}\sum^{\infty}_{N_2=0} z_1^{N_1}z_2^{N_2}Q_{0}(N_1,N_2|N_1^0,N_2^0)
\label{K0def}
\end{equation}
the equation for which is obtained by summing (\ref{Q})
\begin{equation}
a(z_1)\frac{\partial K_0}{\partial z_1}+a(z_2)\frac{\partial K_0}{\partial z_2}+
g(z_1+z_2-2)K_0 = -z_1^{N_1^0}z_2^{N_2^0} 
\label{K0eq}
\end{equation}
with  $a(z_i)=d-(d+b)z_i$.

In order to solve (\ref{K0eq})
we use the method of characteristics (see e.g. \cite{pde:zauderer}).
The characteristic equations are
\begin{eqnarray}
\frac{\D z_1}{\D s} = a(z_1)\label{z1s} \\
\frac{\D z_2}{\D s} = a(z_2)\label{z2s} \\
\frac{\D K_0}{\D s} =
-g(z_1+z_2-2)K_0  -z_1^{N_1^0}z_2^{N_2^0} 
\label{K0s}
\end{eqnarray}
where $s$ is a time-like parameter and (\ref{z1s},\ref{z2s}) define
characteristic  curves along which the 
partial differential equation (\ref{K0eq})
reduces to the ordinary differential equation
(\ref{K0s}).
Solving
(\ref{z1s},\ref{z2s}) yields
the curves 
\begin{equation}
z_1 = \frac{d}{d+b} + A v
\qquad z_2 = \frac{d}{d+b} + B v
\end{equation}
where
\begin{equation}
v =  {\rm e}^{-(d+b)s}
\end{equation}
and $A$,$B$ are two constants to be fixed.

Equation (\ref{K0s}) can be  rewritten as
an ordinary differential equation in $v$ as
\begin{equation}
\fl \frac{\D }{\D v} \left[
K_0(v) v^{\frac{2gb}{(d+b)^2}} {\rm e}^{-\frac{g(A+B)}{(d+b)}v}
\right]
=\frac{1}{d+b}
v^{\frac{2gb}{(d+b)^2}-1}
{\rm e}^{-\frac{g(A+B)}{(d+b)}v}
\left(\frac{d}{d+b} + A v\right)^{N_1^0}
\left(\frac{d}{d+b} + B v\right)^{N_2^0}\;.
 \label{K0de}
\end{equation}
To integrate (\ref{K0de}) we choose  an end-point of the integration
as $v=1$ and set this to correspond to  an arbitrary point
in the $z_1$--$z_2$ plane. This fixes the two constants $A$,$B$
as
\begin{equation}
A= z_1 - \frac{d}{d+b} \qquad B=  z_2 - \frac{d}{d+b}\;.
\end{equation}
Thus we obtain
\begin{eqnarray}
K_0(z_1,z_2)
&=&\frac{1}{d+b}
\int_{0}^1 \D v
v^{\frac{2gb}{(d+b)^2}-1}
\exp \left\{ -\frac{g}{d+b}\left(
\frac{2d}{d+b} -z_1-z_2\right)(1-v)
\right\} \nonumber \\
&&\times \left(\frac{d}{d+b}(1-v)  +  vz_1\right)^{N_1^0}
\left(\frac{d}{d+b}(1-v) + v z_2\right)^{N_2^0}
\;.
\end{eqnarray}
We now expand as a power series in $z_1$, $z_2$ to obtain
$Q_0(N_1,N_2|N_1^0,N_2^0)$ from (\ref{K0def})
\begin{eqnarray}
\fl Q_0(N_1,N_2|N_1^0,N_2^0)
&=
\frac{{\rm e}^{ -\frac{2gd}{(d+b)^2}}}{d+b}
\sum_{p=0}^{N_1} \frac{1}{p!}\left( \frac{g}{d+b}\right)^p
\left( \begin{array}{c} N_1^0 \\N_1-p \end{array} \right)
\left(\frac{d}{d+b}\right)^{N_1^0-N_1+p}
  \label{Q0} \\
&\times  \sum_{r=0}^{N_2} \frac{1}{r!}\left( \frac{g}{d+b}\right)^r
\binom{N_2^0}{N_2-r}
\left(\frac{d}{d+b}\right)^{N_2^0-N_2+r} \nonumber \\
&\times 
I\left(\frac{2gb}{(d+b)^2}{+}N_1{-}p{+}N_2{-}r,
N_1^0{+}N_2^0{-}N_1{-}N_2{+}2p{+}2r{+}1;
\frac{2gd}{(d+b)^2}
\right)\nonumber
\end{eqnarray}
where $I(\alpha,\beta;x)$ is defined as the integral
\begin{equation}
I(\alpha,\beta;x)
=\int_0^1 \D v\,  v^{\alpha-1}(1-v)^{\beta-1} {\e}^{x v}
= \frac{\Gamma(\alpha)\Gamma(\beta)}{\Gamma(\alpha+\beta)}\
{}_1F_1(\alpha,\alpha+\beta;x) \;,
\label{Idef}
\end{equation}
and  $\Gamma(z)$ and  ${}_1F_1(a,b;z)$
are the usual Euler Gamma function and confluent hypergeometric  function
respectively.
\subsection{Green function for $\mathcal{L}_1$}
\label{sec:GL1}
If we define  a Green function for
the  operator $\mathcal{L}_1$ through
\begin{equation}
\mathcal{L}_1 Q_1(N_1,N_2|N^0_1,N^0_2)  =-\delta_{N_1,N_1^0}\delta_{N_2,N_2^0}\;,
\label{Q1}
\end{equation}
this equation only has a solution when $N_2\neq0$.
To see this we note that 
summing the left hand side of (\ref{Q1}) over all $N_1$,$N_2$ yields zero
i.e. the operator conserves probability.
Therefore one cannot solve (\ref{Q1}) or indeed (\ref{LPf}) for an arbitrary
right hand side; one requires that the sum of the right hand side
of (\ref{LPf})  over all
$N_1$,$N_2$ yields zero.
However, since 
the null space of the operator
$\mathcal{L}_1$ is concentrated on $N_2=0$ 
(i.e. the stationary state  in (\ref{P10}) is proportional to $\delta_{N_2,0}$)
we can find a solution of (\ref{Q1})
for  $N_2 > 0$.

It is simplest to proceed by considering
the solution $P_1$ for  an arbitrary right hand side
$-h(N_1,N_2)$
\begin{equation}
\mathcal{L}_1 P_1(N_1,N_2)  = -h (N_1,N_2)
\label{LPh}
\end{equation}
that satisfies
\begin{equation}
\sum_{N_1=0}^{\infty}\sum_{N_2=0}^{\infty}h(N_1,N_2) =0\;.
\label{hcon}
\end{equation}
We define  generating functions
\begin{eqnarray}
K_1(z_1,z_2) &=& 
\sum^{\infty}_{N_1=0}\sum^{\infty}_{N_2=0} z_1^{N_1}z_2^{N_2}P_1(N_1,N_2)\\
H(z_1,z_2) &=& 
\sum^{\infty}_{N_1=0}\sum^{\infty}_{N_2=0} z_1^{N_1}z_2^{N_2}h(N_1,N_2)
\end{eqnarray}
then summing (\ref{LPh}) yields
\begin{equation}
d(1-z_1)\frac{\partial K_0}{\partial z_1}
+d(1-z_2)\frac{\partial K_0}{\partial z_2}+
g(z_1-1)K_0 = -H(z_1,z_2)\;.
\label{K1eq}
\end{equation}

In order to solve (\ref{K1eq})
we again use the method of characteristics.
The characteristic equations are this time
\begin{eqnarray}
\frac{\D z_1}{\D s} = d(1-z_1) \\
\frac{\D z_2}{\D s} = d(1-z_2) \\
\frac{\D K_1}{\D s} =
-g(z_1-1)K_1  - H(z_1,z_2) \;.
\label{K1s}
\end{eqnarray}
Solving the first two equations yields
\begin{equation}
z_1 = 1 + A v
\qquad z_2 = 1  + B v
\end{equation}
where now
\begin{equation}
v =  {\rm e}^{-ds}
\end{equation}
and $A$, $B$ are constants to be fixed.
The  final equation becomes
\begin{equation}
\frac{\D }{\D v}\left[ K_1 {\rm e}^{-A\frac{g}{d}v}\right]
 = \frac{H(z_1,z_2)}{d v} {\rm e}^{-A\frac{g}{d}v}
\label{K1s2}
\end{equation}
We choose the integration to be from $v=0$ to $v=1$
where $v=0$ corresponds to $z_1=z_2=1$ and $v=1$ corresponds to
an arbitrary point in the $z_1$--$z_2$ plane
which implies  $A= z_1-1$ and $B=z_2-1$.
We then obtain the solution of (\ref{K1s2})
\begin{equation}
K_1(z_1,z_2) =
K_1(1,1) {\rm e}^{(z_1-1)\frac{g}{d}}
+\frac{1}{d}
\int_0^1 \D v \frac{{\rm e}^{(z_1-1)\frac{g}{d}(1-v)}}{v}
H(1+(z_1-1)v, 1+(z_2-1)v)\;.
\end{equation}
Expanding as a power series in $z_1$,$z_2$ implies
\begin{eqnarray}
P_1(N_1,N_2)
&=&  K(1,1) {\rm e}^{-g/d}
\frac{(g/d)^{N_1}}{N_1!} \delta_{N_2,0}
\nonumber \\
&+&  {\rm e}^{-g/d} \sum_{p=0}^\infty\sum_{q=0}^\infty 
\frac{h(p,q)}{d} \binom{q}{N_2}
\sum_{r=0}^{N_1}
\left(\frac{g}{d}\right)^{r}
\frac{1}{r!} \binom{p}{N_1-r}
\nonumber \\ &&\times
I(N_1+N_2-r,p+q-N_1-N_2+2r+1; g/d)\;.
\label{P1h}
\end{eqnarray}
As discussed above, for $N_2>0$ we can define the Green function
(\ref{Q1}) 
by means of which the solution of (\ref{LPh}) may be written
\begin{equation}
P_1(N_1,N_2) = \sum_{N_1^0,N_2^0}
Q_1(N_1,N_2|N_1^0,N_2^0)h(N_1^0,N_2^0)\;.
\end{equation}
Then we can read off the Green function
from (\ref{P1h}) as
\begin{eqnarray}
Q_1(N_1,N_2|N_1^0,N_2^0)
&=&  
{\rm e}^{-g/d} 
\frac{1}{d}\binom{N_2^0}{N_2}
\sum_{r=0}^{N_1}
\left(\frac{g}{d}\right)^{r}
\frac{1}{r!} \binom{N_1^0}{N_1-r}\\ &&\times
I(N_1+N_2-r,N_1^0+N_2^0-N_1-N_2+2r+1; g/d)
\nonumber
\end{eqnarray}
For $N_2=0$
the solution (\ref{P1h}) reads
\begin{eqnarray}
P_1(N_1,0)
&=&  K(1,1) {\rm e}^{-g/d}
\frac{(g/d)^{N_1}}{N_1!} 
\nonumber \\
&+&  {\rm e}^{-g/d} \sum_{p=0}^\infty\sum_{q=0}^\infty 
\frac{h(p,q)}{d} 
\sum_{r=0}^{N_1}
\left(\frac{g}{d}\right)^{r}
\frac{1}{r!} \binom{p}{N_1-r}
\nonumber \\ &&\times
I(N_1-r,p+q-N_1+2r+1; g/d)\;.
\label{P1hN20}
\end{eqnarray}
Some  care is required with the $r=N_1$ term
of the sum in (\ref{P1hN20})
since the integral
$I(0,\beta;x)$ does not converge.
However 
the property (\ref{hcon}) of  the function $h$  implies that the coefficient of the
offending integral is zero. To see this
one
can  write the $r=N_1$  term of (\ref{P1hN20}) as
\begin{eqnarray}
{\rm e}^{-g/d} \sum_{p=0}^\infty\sum_{q=0}^\infty 
\frac{h(p,q)}{d} 
\left(\frac{g}{d}\right)^{N_1}
\frac{1}{N_1!}
I(0,p{+}q{+}N_1{+}1;g/d)
\nonumber \\
= \sum_{p=0}^\infty\sum_{q=0}^\infty 
\frac{h(p,q)}{d} 
\left(\frac{g}{d}\right)^{N_1}
\frac{1}{N_1!} 
\int_0^1 \D v\,  {\rm e}^{-\frac{g}{d}(1-v)}v^{-1}(1-v)^{p+q+N_1} \nonumber \\
= \sum_{p=0}^\infty\sum_{q=0}^\infty 
\frac{h(p,q)}{d} 
\left(\frac{g}{d}\right)^{N_1}
\frac{1}{N_1!} 
 \sum_{s=1}^{p+q} 
(-1)^s {\rm e}^{-\frac{g}{d}}
\binom{p+q}{s}
I(s,N_1+1; g/d)\;.
\label{rN1term}
\end{eqnarray}
In the final equality,
the binomial expansion of $(1-v)^{p+q}$ has been
used with the term $s=0$
not present since its coefficient vanishes due to
(\ref{hcon}). All the  integrals in 
 (\ref{rN1term}) then converge.

\subsection{First-order results}
The first-order contribution
to the stationary probability  $P_0^{(1)}(N_1,N_2)$  is given by:
\begin{equation}
 P_0^{(1)}(N_1,N_2) 
=\sum_{N_1^0,N_2^0}f_0^{(1)}(N_1^0,N_2^0)Q_0(N_1,N_2|N_1^0,N_2^0)
\label{P01}
\end{equation}
with 
\begin{equation}
 f_0^{(1)}(N_1^0,N_2^0) =\frac12\exp\left(-\frac{g}{d}\right)
\left[ \left(\frac{g}{d}\right)^{N_1^0-1}\frac{\delta_{N_2^0,0}}{(N_1^0-1)!}+\left(\frac{g}{d}\right)^{N_2^0-1}\frac{\delta_{N_1^0,0}}{(N_2^0-1)!}\right]\;.
\end{equation}

Although $Q_0(N_1,N_2|N_1^0,N_2^0)$ is expressed in terms of known integrals
in (\ref{Q0}), it is advisable to go back to the explicit expression of the integrals (\ref{Idef}) to evaluate the sums
appearing in (\ref{P01}) more easily. 
In that way, $ P_0^{(1)}(N_1,N_2)$ can be written as:
\begin{eqnarray}
\fl  P_0^{(1)}(N_1,N_2) 
=&\frac12\frac{{\rm e}^{-\frac{g}{d} -\frac{2gd}{(d+b)^2}}}{d+b}
\sum_{N_1^0=0}^{\infty}\left(\frac{g}{d}\right)^{N_1^0-1}\frac{1}{(N_1^0-1)!}\sum_{p=0}^{N_1}\frac{1}{p!}\left(\frac{g}{d+b}\right)^{p}\binom{N_1^0}{N_1-p}\left(\frac{d}{d+b}\right)^{N_1^0-N_1+p}\nonumber \\
\times&\frac{1}{N_2!}\left(\frac{g}{d+b}\right)^{N_2}\int_0^1 dv\,   v^{\frac{2gb}{(d+b)^2}+N_1-p-1}(1-v)^{N_1^0-N_1+N_2+2p}e^{\frac{2gd}{(d+b)^2}v}+symm
\label{P01int}
\end{eqnarray} 
where the label {\it symm} refers to the fact that there will be another term equal to the written one, apart from a switch in the variables $N_1^0$ and $N_2^0$.

In  \ref{appB}  it is shown how the expression may be simplified to
the result
\begin{eqnarray}
\fl  P_0^{(1)}(N_1,N_2) 
= &\frac12\exp\left(\frac{g(b-d)}{(d+b)^2}-\frac{g}{d}\right)\left(\frac{g}{d}\right)^{-1}\frac{1}{d+b}\frac{1}{N_2!}\left(\frac{g}{d+b}\right)^{N_2}\sum_{m=0}^{N_1}\left(\frac{g}{d+b}\right)^{N_1-m}\left(\frac{g}{d}\right)^m 
\nonumber \\
\times&\frac{1}{(N_1-m)!m!}\Bigg[\frac{g}{d+b}I\left(\frac{2gb}{(d+b)^2}+m,N_2+N_1-m+2;\frac{g(d-b)}{(d+b)^2}\right) 
\nonumber \\
& +mI\left(\frac{2gb}{(d+b)^2}+m,N_2+N_1-m+1;\frac{g(d-b)}{(d+b)^2}\right)\Bigg]+symm\;.
\label{P01res}
\end{eqnarray}


Once we have $P_0^{(1)}$, we can plug this result into the $P_1^{(1)}$ equation
which becomes for $N_2>0$
\begin{equation}
 P_1^{(1)}(N_1,N_2) 
=\sum_{N_1^0,N_2^0}f_1^{(1)}(N_1^0,N_2^0)Q_1(N_1,N_2|N_1^0,N_2^0)
\end{equation}
where $f_1^{(1)}(N_1,N_2)=-P_1^{(0)}(N_1,N_2)+b(N_1+1)P_0^{(1)}(N_1+1,N_2)$.

We consider  separately the two
terms of $f_1^{(1)}$. The first  is
\begin{equation}
 -P_1^{(0)}(N_1^0,N_2^0)=-\frac12\frac{\exp{-\frac{g}{d}}}{N_1^0!}\left(\frac{g}{d}\right)^{N_1^0}\delta_{N_2^0,0}
\end{equation}
Since the calculation with the Green function
(see section~\ref{sec:GL1}) is only for $N_2 \neq 0$ this term does not contribute and the only contribution comes from the
second term in  $f_1^{(1)}$ involving  $P_0^{(1)}$. The resulting expression, 
obtained by substituting $h=b(N_1+1)P_0^{(1)}(N_1+1,N_2)$ in
(\ref{P1h}) with  $P_0^{(1)}$ given by (\ref{P01res}), is
\begin{eqnarray}
\fl
P_1^{(1)}(N_1,N_2) =&\frac{b}{2g} \exp\left(\frac{g(b-d)}{(d+b)^2}-\frac{2g}{d}\right)\sum_{N_1^0=0}^{\infty}\sum_{N_2^0=0}^{\infty}(N_1^0+1)
\Bigg\{ \label{P11res} \\
&\frac{1}{(d+b)}\frac{1}{N_2^0!}\left(\frac{g}{d+b}\right)^{N_2^0}
\sum_{m=0}^{N_1^0+1}\left(\frac{g}{d+b}\right)^{N_1^0+1-m}\left(\frac{g}{d}\right)^m\frac{1}{(N_1^0+1-m)!m!}\times \nonumber \\
& \Bigg[\frac{g}{d+b}I\left(\frac{2gb}{(d+b)^2}+m,N_2^0+N_1^0-m+3;\frac{g(d-b)}{(d+b)^2}\right)\nonumber \\
& +mI\left(\frac{2gb}{(d+b)^2}+m,N_2^0+N_1^0-m+2;\frac{g(d-b)}{(d+b)^2}\right)\Bigg]+symm(N_1^0+1,N_2^0)\Bigg\}\times \nonumber \\
&\binom{N_2^0}{N_2}
\sum_{r=0}^{N_1}
\left(\frac{g}{d}\right)^{r}
\frac{1}{r!} \binom{N_1^0}{N_1-r}I(N_1+N_2-r,N_1^0+N_2^0-N_1-N_2+2r+1; g/d) 
\nonumber
\end{eqnarray}
Where the symmetric term in this case corresponds to the term inside the curly brackets, 
exchanging the places of $N_1^0+1$ and $N_2^0$ coming from the symmetric term in $P_0^0$.
For $N_2=0$ we obtain the result shown in (\ref{P1hN20}) and cannot be simplified further. 

Equations (\ref{P1hN20}), (\ref{P01res}) and (\ref{P11res})
are the main results of this section and give closed form expressions for the 
first-order contributions to the stationary proabilities.
The normalization constant $K(1,1)$ 
appearing in (\ref{P1hN20}) is obtained from the condition:
\begin{equation}
 \sum_{N_1=0}^{\infty}\sum_{N_2=0}^{\infty}
\left[P_0^{(1)}(N_1,N_2)+P_1^{(1)}(N_1,N_2)+P_2^{(1)}(N_1,N_2)\right]=0\;.
\end{equation}

\begin{figure}[htb]
 \centering
 \begin{minipage}[c]{0.7\linewidth}
    \begin{center}
 \includegraphics[width=\linewidth]{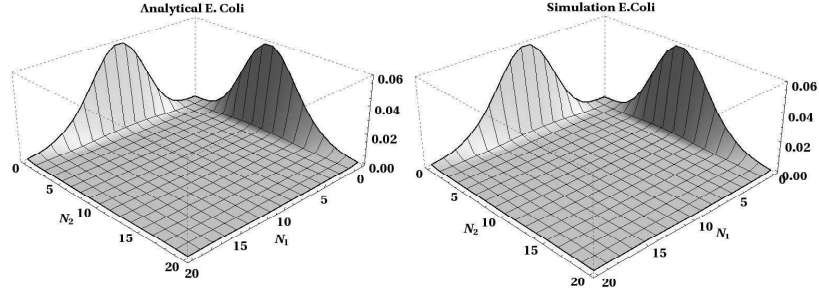}
       \textbf{a}) E. coli values.
    \end{center}
 \end{minipage}
\linebreak
  \begin{minipage}[c]{0.7\linewidth}
    \begin{center}
 \includegraphics[width=\linewidth]{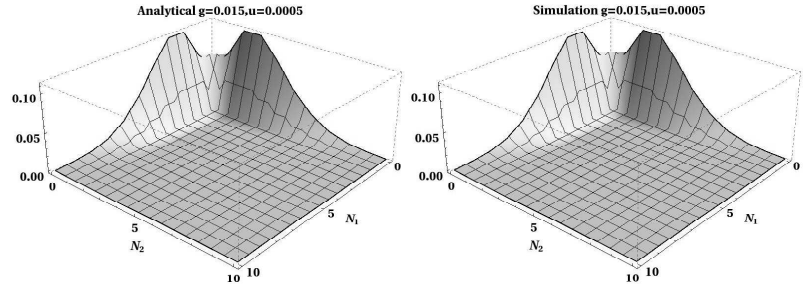}
       \textbf{b}) $g=0.015,u=0.0005,d=0.005,b=0.1$.
    \end{center}
 \end{minipage}
\linebreak
 \begin{minipage}[c]{0.7\linewidth}
    \begin{center}
 \includegraphics[width=\linewidth]{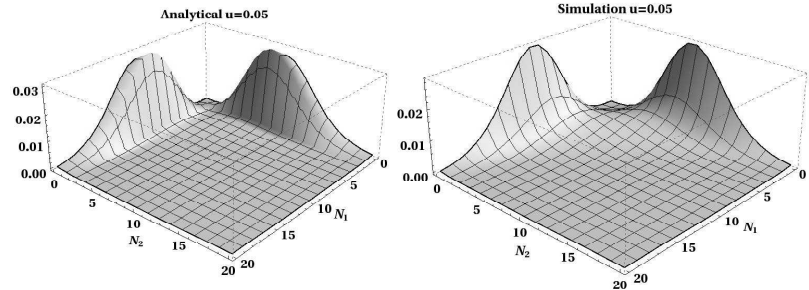}
       \textbf{c}) E. coli values with $u=0.05$.
    \end{center}
 \end{minipage}
\caption{
Comparison between the analytical first order and the simulation distributions for different values of the parameters
a) E. coli values (\ref{ecv}). The probability distributions have the same shape and the order of magnitude is well reproduced in almost every point. 
b) $g=0.015,u=0.0005,d=0.005,b=0.1$ The order of magnitude is again well reproduced. In this case, the two peaks get closer to the origin as the ratio g/d is smaller. c) E. coli values with $u=0.05$. The approximation at first order is 
no longer accurate as the value of $u$ is no longer small compared to the rest of the parameters of the model.}
\label{fig:finalpics}
 \end{figure}

Once the values of the probabilities  $P_i^{(1)}(N_1,N_2)$
  have been obtained numerically, they are multiplied by the
  unbinding parameter $u$ and added to the zeroth order probabilities.
  In that way, the probability distributions $P_i(N_1,N_2)$ can be
  computed and plotted up to the first order of the expansion. Figure
  \ref{fig:finalpics} shows the probability distributions for
  different values. The agreement is visually good in the first two
  examples: typical E. coli values (\ref{ecv}) and another interesting
  case, with smaller $g$. The order of magnitude is well reproduced in
  almost every point of the probability distribution.  However, the
  approximation does not work  accurately if the value of the unbinding
  parameter $u$ is of the order or bigger than the other parameters.

\begin{figure}[htb]
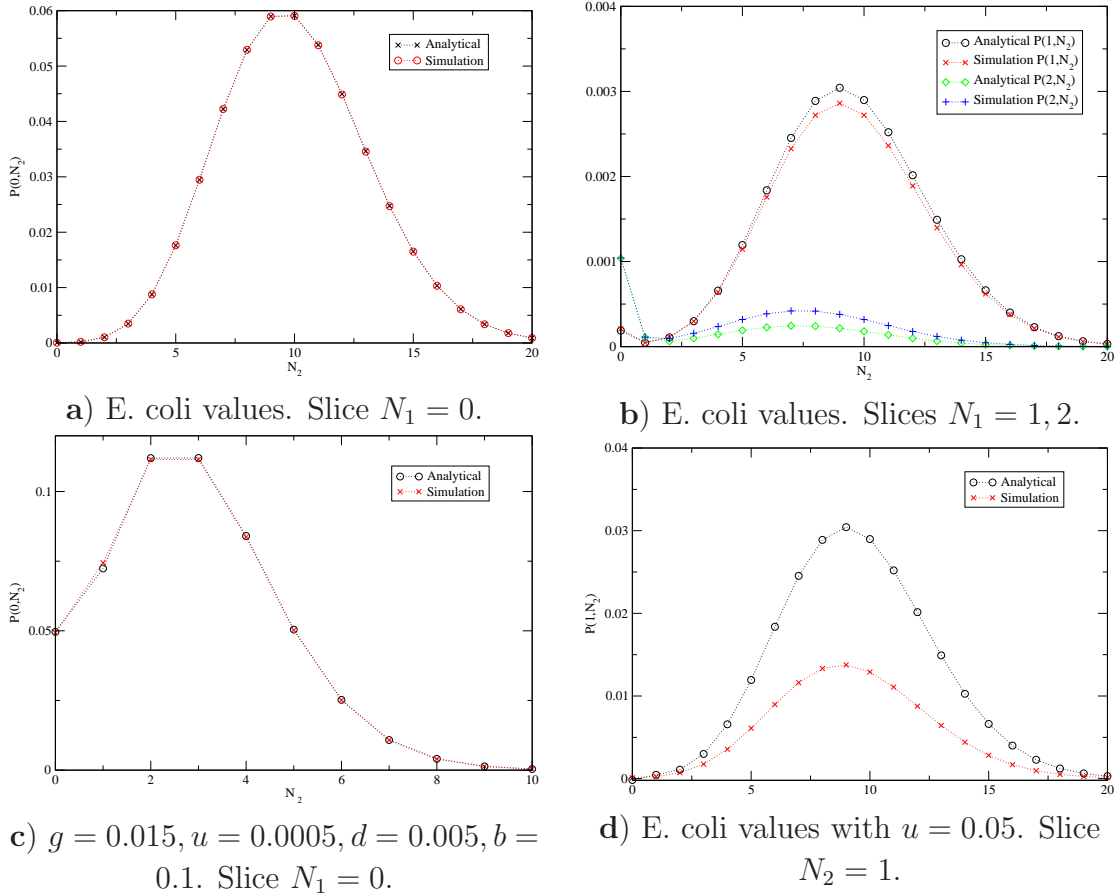

 \centering
 \begin{minipage}[c]{0.45\linewidth}
    \begin{center}
 \includegraphics[width=\linewidth]{ecoliaxis0.eps}
       \textbf{a}) E. coli values. Slice $N_1=0$.
    \end{center}
 \end{minipage}
 \hspace{0.02\linewidth}
\begin{minipage}[c]{0.45\linewidth}
    \begin{center}
 \includegraphics[width=\linewidth]{ecoliaxis12.eps}
       \textbf{b}) E. coli values. Slices $N_1=1,2$.
    \end{center}
 \end{minipage}
\linebreak
\begin{minipage}[c]{0.45\linewidth}
    \begin{center}
 \includegraphics[width=\linewidth]{g0.015u5-4axis0.eps}
       \textbf{c}) $g=0.015,u=0.0005,d=0.005,b=0.1$. Slice $N_1=0$.
    \end{center}
 \end{minipage}
 \hspace{0.02\linewidth}
\begin{minipage}[c]{0.45\linewidth}
    \begin{center}
 \includegraphics[width=\linewidth]{u0.05axis1.eps}
       \textbf{d}) E. coli values with $u=0.05$. Slice $N_2=1$.
    \end{center}
 \end{minipage}
\caption{Comparison between the analytical probability distributions
  $P(N_1^*,N2)$ where $N_1^*$ is fixed and chosen to correspond to
  slices with largest probability mass. The agreement in slice $N_1=0$
  is good for both figures a) and c). There is some quantitative
  difference in b), which corresponds to the E. coli values in slices
  with less probability mass. This is improved with second order
  calculations, as can be seen in figure \ref{fig:axes2}. In d) the
  difference is clear, since the first order approximation is no
  longer accurate, as discussed in figure \ref{fig:finalpics}. The
  error in the values is negligible and, in all the cases, is smaller
  than the size of the used symbols.}
\label{fig:axes1}
 \end{figure}

Although the first two examples of figure \ref{fig:finalpics} seem visually in good agreement with the simulations, the 
best way to check this is to plot different slices of the probability distribution, that is   the probability distribution of $N_2$ where
$N_1$ is held constant (figure \ref{fig:axes1}).
We choose the values of $N_1$ to correspond to the slices with large probability mass, i.e. $N_1 = 0,1,2$ for E. coli values.
Figure \ref{fig:axes1}) shows that along the slice with greatest probability mass ($N_1=0$)
the analytical and simulation plots  show good agreement.
For $N_1>0$, there is reasonable agreement
for the E. coli values  (figure \ref{fig:axes1} b)),
whereas for larger $u$ 
the agreement is not so good (figure \ref{fig:axes1} d)).

\begin{figure}
\centering
\caption{Comparison between the  probability slices $P(1,N_2)$ and $P(2,N_2)$ from second order analytical calculations and simulations, for E. coli values. The second order is clearly enough to get accurate results.}
\label{fig:axes2}
\includegraphics[scale=0.4]{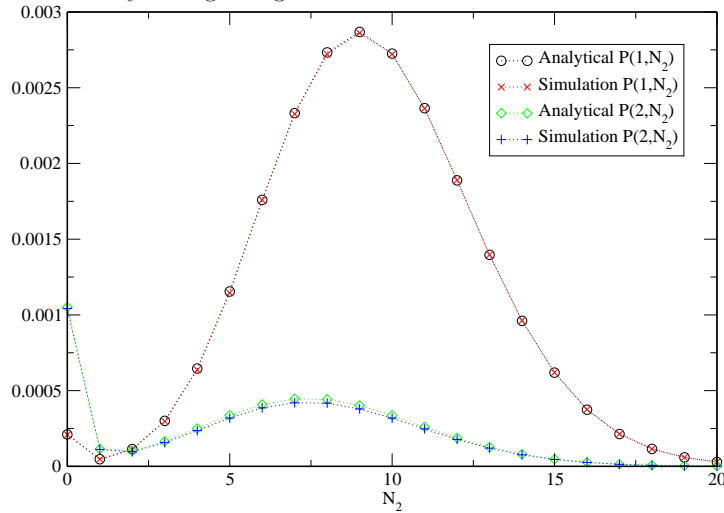}
\end{figure}

Since the proposed method is general, calculations can be performed up
to any necessary order to get better results. For example, in the case
of E. coli in axes $N_1=1$, $N_2=1$ is enough to compute the second
order to have very good results (figure \ref{fig:axes2}). In general,
the method can be iterated as many times as required.

\section{Conclusion}
The exclusive genetic switch presented in this paper represents a
minimal model of two populations that compete in an indirect way, that
is, through a common promoter site that
when occupied by a member of one population stops the production of the
other. It is a non-equilibrium system because the microscopic
processes are irreversible and detailed balance does not hold.  The
non equilibrium stationary state exhibits interesting properties of
bistability i.e. generically at any given time the system is dominated
by one of the populations.  We have developed two main analytical
approaches to study the stationary state: a mean field theory and an
exact perturbative approach. In addition we have presented
some exactly solvable limits. We also have studied the system and
checked the analytical results by using Monte Carlo simulations.

These simulations show that the symmetry of the system is always
broken, that is, the system is always functioning as a switch, with
two opposite states in which one population is much more abundant than
the other. This holds
generically except in the limit $u,b\to \infty$ where the state is symmetric.
As the relevant parameters for the switch become smaller,
the probability of finding bound proteins decreases, but the
distributions always remain asymmetric, showing no phase transition
between  asymmetric and symmetric states

The mean field theory of section~\ref{sec:mft} 
is based on exact moment equations which are
factorised at the level of second moments. The approximation scheme is
constructed to be exact in the limiting parameter case of vanishing
binding/unbinding rates $b$, $u$ with the binding constant
$k=\frac{b}{u}$ held constant
The theory
shows good agreement with some quantities obtained from the
simulation, but there are some other quantities whose agreement varies
and depends on the value of the parameters.  The mean-field theory
could be improved upon by systematically considering higher order
moments and correlations.

As a first attempt, to our knowledge, at an exact solution of this nonequilibrium system
we have developed an exact perturbative approach which
consists of an expansion in the unbinding rate $u$.
We computed the expansion to first order and this has allowed
  us to obtain the whole probability distribution for the typical
  E. coli and other values, in good agreement with the
  simulations. Higher orders can be obtained systematically by
  iterating the proposed method.  The analytical expressions at first
  order (\ref{P01res},\ref{P11res}) already illustrate the complexity
  of the nonequilibrium stationary state.  In principle, the Green
  functions that have been calculated in Section 4 allow the expansion
  to be carried out to arbitrary order although analytical expressions
  for the higher order terms in the expansions might be long or
  difficult to simplify. However, they still can be computed
  numerically by programming the proposed operations and, in
  principle, the method can be iterated as many times as necessary to
  get more accurate results.

In this paper we have not attempted to study the dynamics, but it
would be of interest to do so. In particular the dependence of the
flip time or first passage time (the time for the system to change
from being dominated by one population to the other) on the parameter
values is of interest. It may be possible to extend our analytical
solution to consider such flip times and to provide estimate of first
passage times between the two bistable states.  It might be that the
Green functions we have computed in section 4 contain some useful
dynamical information.

The techniques that we have developed should be applicable to the
understanding of other related systems and properties of genetic
switches in general.  The factorisation of the moment equation
hierarchy is straightforward to implement to obtain a mean field
theory; the exact perturbative approach is more involved but can be
used in general for problems with similar probability distributions,
as long as the Green functions are analytically solvable.  Thus the
techniques represent standard procedures to analyze this class of
systems.

\vspace{2em}
\section*{Acknowledgements}
Juan Venegas-Ortiz would like to acknowledge the award of a College
Studentship from the University of Edinburgh, and would like to thank
Francisco Cordob\'es-Aguilar for useful discussions about numerical
implementation of the method.

\appendix
\section{Analytical method for systems of linear PDEs.}
According to \cite{pde:zauderer}, there is a way to solve, or at least simplify, a system of linear first-order partial differential equations. First of all, considering a three dimensional space with a vector $\vec{K}(z_1,z_2)$ for the three probabilites, the system has to be written as:\begin{equation}
A(z_1,z_2)\frac{\partial\vec{K}(z_1,z_2)}{\partial z_1}+B(z_1,z_2)\frac{\partial\vec{K}(z_1,z_2)}{\partial z_2}=C(z_1,z_2)\vec{K}(z_1,z_2)+\vec{d}(z_1,z_2)\end{equation}where $A$,$B$ and $C$ are $3\times 3$ matrices, and $\vec{d}$ and $\vec{K}$ are column matrices. The requirement to solve the system is that $det A(z_1,z_2) \neq 0$ or $det B(z_1,z_2) \neq 0$.

In this case the expression of the matrices is:
\begin{eqnarray}
A(z_1,z_2) =
\left(
\begin{array}{ccc} d-(d+b)z_1 & 0 & 0\\
b & d(1-z_1) & 0\\0 & 0 & d(1-z_1)\\
\end{array}
\right)\nonumber \\
\nonumber \\
B(z_1,z_2) =
\left(
\begin{array}{ccc} d-(d+b)z_2 & 0 & 0\\
0 & d(1-z_2) & 0\\b & 0 & d(1-z_2)\\
\end{array}
\right)\nonumber \\
\nonumber \\
C(z_1,z_2) =
\left(
\begin{array}{ccc} -g(z_1+z_2-2) & -uz_1 & -uz_2\\
0 & -[g(z_1-1)-u] & 0\\0 & 0 & -[g(z_2-1)-u]\\
\end{array}
\right)\nonumber \\
\nonumber \\
d(z_1,z_2) =
\left(
\begin{array}{c}  0\\
 0\\0
\end{array}
\right)
\end{eqnarray}
This is not the most general system that could be written with this notation, since the matrix $\vec{d}$ is zero, and $A$ and $B$ only depend on $z_1$, $z_2$, respectively.

The requirement to solve the system is that $det A(z_1,z_2) \neq 0$ or $det B(z_1,z_2) \neq 0$, which is fulfilled almost in every point. Multiplying by the matrix $B^{-1}$, the matrix $A$ is transformed into: 
\begin{equation}
\begin{eqalign}
A'=B^{-1}A =
\left(
\begin{array}{ccc}
\frac{d-(d+b)z_1}{d-(d+b)z_2} & 0 & 0\\
\frac{b}{d(1-z_2)} & \frac{(1-z_1)}{1-z_2} & 0\\\frac{-b[d-(d+b)z_1]}{d(1-z_2)(d-(d+b)z_2)} & 0 & \frac{(1-z_1)}{1-z_2}\\
\end{array}
\right)
\\ 
\end{eqalign}
\end{equation}
whose eigenvalues, written as columns in a matrix $R$ are:
\begin{equation}
R =
\left(
\begin{array}{ccc}
 0 & 0 & \frac{-d(z_1-z_2)}{d(z_1-1)+bz_1}\\
0 & 1 & \frac{d-(d+b)z_2}{d-(d+b)z_1}\\1 & 0 & 1
\end{array}
\right)\end{equation}
This hyperbolic system can be transformed with elementary matrix operations:
\begin{equation} 
\begin{eqalign}
C' =B^{-1}C 
\\
\vec{K} =R(z_1,z_2)\vec{v}
 \end{eqalign}
\end{equation}
The method then states that the components of $v$ obey the following system of equations:
\begin{equation}
\frac{dv_i}{dz_2} =\sum_{j=0}^{2} \hat{c}_{ij}v_j
\end{equation}
where $\hat{c}_{ij}$ are the components of the matrix $C'$ once we have performed the transformation with $R$: $\hat{C}=R^{-1}C'R$.

As can be seen from the last equation, we have uncoupled the derivative terms and, even if this system cannot be solved analytically, it is easier to deal with it computationally. The problem itself can be written as:
\begin{eqnarray}
\frac{dv_0}{dz_2} &=&
\left[\frac{(g+u)z_1-gz_2}{d(z_1-z_2)}\right]v_0
+\left[\frac{uz_1(z_1-1)}{d(z_1-z_2)(z_2-1)}\right]v_1\nonumber\\  
&&+\left[\frac{-dg(z_1-1)^2+b[d-(g+u)z_1+gz_1z_2]}{d(z_2-1)[d(z_1-1)+bz_1]}\right]v_2\nonumber \\on&& \quad \frac{dz_1}{dz_2}=\frac{1-z_1}{1-z_2}\nonumber\\
\frac{dv_1}{dz_2} &=&\left[\frac{uz_2}{d(z_2-z_1)}\right]v_0+\left[\frac{(z_1-1)[-gz_1+(g+u)z_2]}{d(z_2-z_1)(z_2-1)}\right]v_1\nonumber\\
&&+\left[\frac{dg(z_2-1)^2+b[-d+(g+u)z_2-gz_1z_2]}{d(z_2-1)[d(z_1-1)+bz_1]}\right]v_2\nonumber \\
on && \quad \frac{dz_1}{dz_2}=\frac{1-z_1}{1-z_2}\nonumber \\
\frac{dv_2}{dz_2} &=&\left[\frac{uz_1}{d(z_2-z_1)}+\frac{u}{d-(d+b)z_2}\right]v_0+\left[\frac{uz_1[d(z_1-1)+bz_1]}{d(z_1-z_2)(d-(b+d)z_2)}\right]v_1
\nonumber \\
&+&\left[\frac{-u+g(z_1+z_2-2)}{d(z_2-1)[d(z_2-1)+bz_2]}\right]v_2
\nonumber \\
on&& \quad \frac{dz_1}{dz_2}=\frac{d-(d+b)z_1}{d-(d+b)z_2} 
\label{exactsol}
\end{eqnarray}
Although this method in principle
solves exactly the system of partial differential equations
it appears a formidable task to actually integrate equations
(\ref{exactsol}).\\

\section{Derivation of expression (\ref{P01res})}
\label{appB}
In this appendix we give the detailed
derivation of (\ref{P01res}). We begin from
(\ref{P01int})
\begin{eqnarray}
\fl  P_0^{(1)}(N_1,N_2) 
=&\frac12\frac{{\rm e}^{-\frac{g}{d} -\frac{2gd}{(d+b)^2}}}{d+b}
\sum_{N_1^0=0}^{\infty}\left(\frac{g}{d}\right)^{N_1^0-1}\frac{1}{(N_1^0-1)!}\sum_{p=0}^{N_1}\frac{1}{p!}\left(\frac{g}{d+b}\right)^{p}\binom{N_1^0}{N_1-p}\left(\frac{d}{d+b}\right)^{N_1^0-N_1+p}\nonumber \\
\times&\frac{1}{N_2!}\left(\frac{g}{d+b}\right)^{N_2}\int_0^1 dv\,   v^{\frac{2gb}{(d+b)^2}+N_1-p-1}(1-v)^{N_1^0-N_1+N_2+2p}e^{\frac{2gd}{(d+b)^2}v}+symm
\end{eqnarray} 
where the label {\it symm} refers to the fact that there will be another term equal to the written one, apart from a switch in the variables $N_1^0$ and $N_2^0$.

Now defining:
\begin{equation}
 c(v)\equiv v^{\frac{gb}{(d+b)^2}}e^{-\frac{gd}{(d+b)^2}(1-v)}, \quad m\equiv N_1-p, \quad r(v)\equiv\frac{d}{d+b}\frac{1-v}{v}, \quad s(v)\equiv \frac{g}{d+b}(1-v)
\end{equation}
we arrive at the expression (for convenience we will drop the dependence of the previous functions on $v$):
\begin{equation}
\eqalign{
P_0^{(1)}(N_1,N_2) =&\int_0^1\frac{dv}{v(d+b)}\sum_{N_1^0=0}^{\infty}\frac12
\exp{\left(-\frac{g}{d}\right)}\left(\frac{g}{d}\right)^{N_1^0-1}\frac{c^2}{(N_1^0-1)!} \cr
 & \sum_{m=0}^{N_1}\binom{N_1^0}{m}r^{N_1^0-m}v^{N_1^0}\times\frac{1}{(N_1-m)!}s^{N_1-m}\frac{s^{N_2}}{(N_2)!}+symm
}
\end{equation}

Separating the parts of the expression that can be summed, the following simplification can be obtained by changing the order of the sums appropriately:
\begin{equation}
\eqalign{
  \sum_{N_1^0=0}^{\infty}\left(\frac{g}{d}\right)^{N_1^0-1}\frac{1}{(N_1^0-1)!}v^{N_1^0}\sum_{m=0}^{N_1}\binom{N_1^0}{m}\frac{r^{N_1^0-m}s^{N_1-m}}{(N_1-m)!} \cr
=\left(\frac{g}{d}\right)^{-1}\sum_{m=0}^{N_1}\frac{s^{N_1-m}}{(N_1-m)!}\sum_{N_1^0=m}^{\infty}\binom{N_1^0}{m}\frac{r^{N_1^0-m}}{(N_1^0-1)!}\omega^{N_1^0} \cr
=\left(\frac{g}{d}\right)^{-1}\sum_{m=0}^{N_1}\frac{s^{N_1-m}}{(N_1-m)!m!}\sum_{n=0}^{\infty}\frac{(n+m)}{n!}r^{n}\omega^{m+m} \cr
=\left(\frac{g}{d}\right)^{-1}\sum_{m=0}^{N_1}\frac{s^{N_1-m}}{(N_1-m)!m!}\omega^m\sum_{n=0}^{\infty}\left(\frac{(r\omega)^{n}}{(n-1)!}+\frac{m(r\omega)^{n}}{n!}\right)
\cr
=\left(\frac{g}{d}\right)^{-1}\sum_{m=0}^{N_1}\frac{s^{N_1-m}}{(N_1-m)!m!}\omega^m(s+m)e^s
}
\end{equation}
where $\omega = vg/d$.
Note that this sum could be simplified further, but that the
simplification will not allow us to perform the integration over
$v$ in closed form. In this
and following equation, we will try to obtain  the simplest expressions
globally, knowing that simplifying one part can lead to further
complications in another.

Plugging this sum into the $P_0^{(1)}$ equation and writing all the
explicit forms of the functions, we obtain
the result (\ref{P01res}).



\end{document}